\documentstyle[psfig]{mn}

\def\etal{{\it et al.\ }}
\def\eg{{\it e.g.\ }}

\def\spose#1{\hbox to 0pt{#1\hss}}
\def\approxlt{\mathrel{\spose{\lower 3pt\hbox{$\sim$}}
	\raise 2.0pt\hbox{$<$}}}
\def\approxgt{\mathrel{\spose{\lower 3pt\hbox{$\sim$}}
	\raise 2.0pt\hbox{$>$}}}
\def\approxpropto{\mathrel{\spose{\lower 3pt\hbox{$\sim$}}
	\raise 2.0pt\hbox{$\propto$}}}
\mathchardef\twiddle="2218

\def\multleft#1{\hbox to size{\vbox {\halign {\lft{##}\cr #1}}\hfill}\par}
\def\multright#1{\hbox to size{\vbox {\halign {\rt{##}\cr #1}}\hfill}\par}
\def\Mdot{\hbox{$\dot M$}}
\def\mdot{\hbox{$\dot m$}}

\def\today{\ifcase\month\or January\or February\or March\or April\or May\or
      June\or July\or August\or September\or October\or November\or December\fi
      \space\number\day, \number\year}
\def\<{\thinspace}

\newcommand{\beq}{\begin{equation}}
\newcommand{\eeq}{\end{equation}}
\def\lsim{\mathrel{\mathpalette\@versim<}}
\def\gsim{\mathrel{\mathpalette\@versim>}}
\def\ro{r_{\rm out}}

\def\erg{{\rm\thinspace erg}}

\def\g{{\rm\thinspace g}}

\def\K{{\rm\thinspace K}}
\def\keV{{\rm\thinspace keV}}

\def\km{{\rm\thinspace km}}
\def\kpc{{\rm\thinspace kpc}}

\def\Mpc{{\rm\thinspace Mpc}}
\def\Msun{\hbox{$\rm\thinspace M_{\odot}$}}

\def\s{{\rm\thinspace s}}
\def\yr{{\rm\thinspace yr}}

\def\ergps{\hbox{$\erg\s^{-1}\,$}}

\def\gps{\hbox{$\g\s^{-1}\,$}}
\def\kmps{\hbox{$\km\s^{-1}\,$}}

\def\Msunpyr{\hbox{$\Msun\yr^{-1}\,$}}

\def\kmpspMpc{\hbox{$\kmps\Mpc^{-1}$}}

\title{Low radiative efficiency accretion in the nuclei of elliptical galaxies} 
\author[Low radiative efficiency accretion in the nuclei of elliptical galaxies]
{\parbox[]{6.in} {T.Di Matteo$^1$\thanks{Chandra Fellow}, E.Quataert$^1$, S.W.Allen$^2$, R.Narayan$^1$ and A.C.Fabian$^2$\\
\footnotesize
1. Harvard-Smithsonian Center for Astrophysics, 60 Garden Street, Cambridge MA 02138 \\
2. Institute of Astronomy, Madingley Road, Cambridge CB3 OHA\\
}}
\begin{document}
\maketitle

\begin{abstract}
\noindent The discovery of hard, power-law X-ray emission (Paper I)
from a sample of six nearby elliptical galaxies, including the
dominant galaxies of the Virgo, Fornax and Centaurus clusters (M87,
NGC 1399 and NGC 4696, respectively), and NGC 4472, 4636 and 4649 in
the Virgo cluster, has important implications for the study of
quiescent supermassive black holes. We describe how the broad band
spectral energy distributions of these galaxies, which accrete from
their hot gaseous halos at rates comparable to their Bondi rates, can
be explained by low-radiative efficiency accretion flows in which a
significant fraction of the mass, angular momentum and energy is
removed from the flows by winds.  The observed suppression of the
synchrotron component in the radio band (excluding the case of M87)
and the systematically hard X-ray spectra, which are interpreted as
thermal bremsstrahlung emission, support the conjecture that
significant mass outflow is a natural consequence of systems accreting
at low-radiative efficiencies.  We briefly discuss an alternative
model for the observed X-ray emission, namely that it is due to
nonthermal synchrotron-self Compton processes in the accretion flow,
or wind.  This appears to require implausibly weak magnetic
fields. Emission from a collimated jet viewed off axis should be
distinguishable from the bremsstrahlung model by variability and
thermal line emission studies. We argue that the difference in
radiative efficiency between the nuclei of spiral and elliptical
galaxies arises from the different manner in which interstellar gas is
fed into the nuclei. In ellipticals, matter fed from the hot (slowly
cooling) ISM is likely to be highly magnetized and with low specific
angular momentum, both of which favor low-radiative efficiency
accretion solutions and possibly the formation of the observed jets.
  

\end{abstract}
\begin{keywords}
galaxies: individual: M87 -- NGC 1399 -- NGC 4696 -- NGC 4649 -- NGC 4472 -- NGC 4636, galaxies: active, accretion, accretion flows clusters: general -- cooling flows -- intergalactic medium -- 
X-rays: galaxies 
\end{keywords}

\section{Introduction}
 
The discovery of hard, power-law, X-ray emission associated with
supermassive black holes in the nuclei of six nearby elliptical
galaxies (Allen, Di Matteo \& Fabian 1999; hereafter Paper I) has
brought into sharper focus the issue of quiescent accretion and the
potential ubiquity of low-level nuclear activity in early-type
galaxies.

The nuclei of elliptical galaxies provide excellent environments for
studying the physics of low-luminosity accretion.  There is strong
evidence, from high-resolution optical spectroscopy and photometry,
that black holes with masses of $10^8-10^{10} \Msun$ reside at the
centers of bulge dominated galaxies, with the black hole mass being
roughly proportional to the mass of the stellar component (\eg
Kormendy \& Richstone 1995, Magorrian \etal 1998; van der Marel 1998;
Ho 1998).

X-ray studies of elliptical galaxies also show that they possess
extensive hot gaseous halos that fill their gravitational potentials.
Given the very large inferred black hole masses, this gas must
inevitably be accreting at a rate which can be estimated from Bondi's
(1952) spherical accretion theory (but, see Gruzinov 1999). Such
accretion should, however, give rise to far more nuclear activity (\eg
quasar-like luminosities) than is observed if the radiative efficiency
were as high as 10 per cent (\eg Fabian \& Canizares 1988), as is
generally postulated in standard accretion theory.

Accretion with such high radiative efficiency need not necessarily
occur in nearby ellipticals, however. The scheme proposed by Rees et
al. (1982; see also Begelman 1986; Fabian and Rees 1995) and
successfully applied to a number of giant ellipticals in the Virgo
cluster (M87: Reynolds \etal 1996; NGC 4649: Di Matteo \& Fabian
1997a; Mahadevan 1997; Di Matteo \etal 1999, hereafter DM99) suggests
that the final stages of accretion in elliptical galaxies may occur
via an advection-dominated accretion flow (ADAF; Narayan \& Yi 1995b;
Abramowicz \etal 1995; for recent review see Narayan, Mahadevan \&
Quataert 1998). For such an accretion mode, the quiescence of the
elliptical galaxy nuclei is not surprising; when the accretion rate is
low, the radiative efficiency of the accreting (low density) material
will also be low. Other factors may also contribute to the low
luminosities observed. As discussed theoretically by Blandford \&
Begelman (1999; see also Narayan \& Yi 1994; 1995a) and given
observational credence by DM99, winds/outflows may transport energy,
angular momentum and mass out of the hot, radiatively inefficient
accretion flows, resulting in only a small fraction of the material
supplied at large radii actually accreting onto the central black
hole.

In fact, the central aim of this paper is to argue that the new X-ray
(Paper I; and radio) constraints on nearby ellipticals provide strong
support that a different accretion mode operates in these systems (as
discussed also in Fabian \& Rees 1995); this work also emphasizes the
importance of mass loss in sub-Eddington, radiatively inefficient,
accretion flows.

We present detailed models for the broad-band emission spectra of the
same sample of six giant ellipticals studied in Paper I: M87 (also
known as NGC 4486, the central galaxy of the Virgo Cluster), three
other giant ellipticals in the Virgo Cluster (NGC 4649, NGC 4472 and
NGC 4636; previously studied at high radio and sub-millimeter
frequencies by DM99) and the central galaxies of the Fornax and
Centaurus clusters (NGC 1399 and 4696, respectively). Hard power-law
X-ray emission was recently discovered in all of these galaxies (Paper
I). Most of the sources also show a noticeable lack of high frequency
radio emission (DM99). Together with the absence of strong optical/UV
flux (the ``big blue bump''), or corresponding infrared (IR) flux (Paper I) this
identifies these objects as a new class of accreting sources.

In Section 2, we will argue for low-radiative efficiency in the target
sources by calculating the Bondi accretion rates and associated
luminosities and showing that these latter values are much {\it
greater} than those observed. We indicate also how radio and X-ray
observations are crucial for constraining the properties of low
radiative efficiency accretion. In \S3 and \S4 we summarize the data
and the model calculations with particular emphasis on the effects of
outflows on such models.  Section 5 explicitly compares model
predictions to the data and \S6 provides some discussion of the
implications for our models within the context of elliptical
nuclei. Other possible contributions to the emission are briefly
discussed in \S7 and a prediction of the expected variability
timescales is given in \S8. We conclude and discuss our understanding
of accretion from the hot ISM in elliptical nuclei in \S9.

\section{Clues for understanding low radiative efficiency flows}

\begin{table*}
\caption{Black hole masses (Magorrian et al. 1998), ISM densities,
Bondi rates and radiative efficiencies}
\label{t:bh}
\begin{center}
\begin{tabular}{cccccc}\hline
Object & Black Hole & $ n_{\rm e}$(ISM) & $\zeta$ & $\Mdot_{\rm Bondi}$& $ L_{\rm Bondi}/ L_{\rm Obs}$ \\
\vspace{0.3cm}& ($10^9 \Msun$) & (cm$^{-3}$) at 1~kpc & & \Msunpyr  &    \\
\hline

M87  & 3.0 & 0.24 & 1.02 & 1.5 &$10^{-5}$  \\

NGC 1399 & 5.2 & 0.16 & 1.37 & 3 & $2 \times 10^{-6}$ \\

NGC 4696 & - & 0.40  & 0.93  & - & $ \sim 10^{-5}$\\

NGC 4649 & 3.9 & 0.12 & 1.17 & 1.4 & $3 \times 10^{-5}$ \\

NGC 4472 & 2.6 & 0.10 & 1.27  & 0.7  & $10^{-5}$\\

NGC 4636 & 0.22 & 0.08 & 1.34 & 0.3 &$3 \times 10^{-4}$ \\

\hline
\end{tabular}
\end{center}
\end{table*}

\subsection{Black hole masses and accretion rates}

The objects we consider are excellent examples of quiescent black
holes. The radio galaxy M87 (NGC 4486), which lies at the center of
the Virgo Cluster, has long been known to host a (low-luminosity)
active nucleus, which powers both a relativistic jet and giant radio
lobes. Both M87 and NGC 1399, the dominant galaxy of the Fornax
Cluster, possess central, supermassive black holes with masses
exceeding $10^9 \Msun$ (Ford et al. 1994, Harms et al. 1994, Macchetto
et al. 1997, Magorrian et al. 1998: see Table~\ref{t:bh}).  Although
no direct mass measurement for the central black hole in NGC 4696 has
been made, this is the most luminous galaxy in our sample and is
therefore likely to have the largest black hole. Both NGC 1399 and NGC
4696 also exhibit FR-I type radio activity.

The central cluster galaxies exist in highly gas-rich environments
({\it i.e.} in cooling flows at the centers of their host clusters;
the integrated mass deposition rates from the cooling flows in and
around M87 and NGC 4696 are $\sim 40-50$ \Msunpyr (Allen et al. in
preparation). Although the other three Virgo ellipticals included in
our sample (NGC 4649, NGC 4472 and NGC 4636) do not exist in such
extreme environments, they also have measured black hole masses
(Magorrian et al. 1998; also summarized in Table ~\ref{t:bh}) and
posses hot, rich, (also slowly cooling) X-ray emitting interstellar
media (ISM) to feed the accretion flow.

Hot gas in the cooling flow or the potential well of the galaxy may be
able to smoothly evolve into a hot, radiatively inefficient, accretion
flow after it passes through the ``Bondi accretion radius,'' defined
to be the point where the gravitational potential of the central black
hole begins to dominate the dynamics of the hot gas; the accretion
radius is given by $R_{\rm A} \approx (c/c_s)^2 R_S$, where $c$ is the speed
of light, $c_{\rm s} \sim 10^4T^{1/2}$ cm s$^{-1}$ the sound speed
(and $T$ the gas temperature) and $R_{\rm S} = 2GM/c^2$ the
Schwarzschild radius of the black hole.
As long as the cooling time for the gas is longer than the free fall
time (i.e the ISM density is low enough), the gas is able
to stay hot as it passes through $\sim R_{\rm A}$ and the accretion rate
onto the black hole can be roughly calculated using Bondi's (1952)
formula, which requires an estimate of the density and temperature of
the gas near $\sim R_{\rm A}$.

We have carried out a deprojection analysis of {\it ROSAT} High
Resolution Imager (HRI) observations for our sample of galaxies, with
the primary aim of measuring the central X-ray emitting gas densities
in these systems. The results (quoted at a fixed radius of 1 kpc) are
summarized in Table~\ref{t:bh}. The inferred gas densities are
relatively insensitive to assumptions about the underlying
gravitational potential in the galaxies and host clusters and are
typically characterized by an $R^{-\zeta}$ profile, where $\zeta$ is
measured to be $\sim 1$ (Table~\ref{t:bh}). The measured densities for
the central cluster galaxies are, as expected, slightly higher than
for the Virgo ellipticals (see also Fig. 3, Paper I).

In order to estimate the accretion rates, we extrapolate the observed
density profiles from $1$ kpc to $\approx R_{\rm A}$ using the power-law
models listed in Table~\ref{t:bh}.  For the systems of interest, \beq
R_{\rm A} \sim 0.1 \left(M \over 10^9 M_\odot\right) \left(c_s \over 300
  {\rm km s^{-1}} \right)^{-2} \ {\rm kpc}.
\label{ra} \eeq Thus, {\it ROSAT} observations  at $\approx 1$ kpc probe 
the gas structure reasonably close to $R_{\rm A}$.  

If the influence of the point mass becomes significant at a
temperature $T$ (where the ISM density $\rho_{\rm A} = \rho(1 \kpc) [1
\kpc/R_{\rm A}]^{\zeta}$), the accretion rate is roughly given by
\begin{equation} 
\Mdot = 4\pi R_{\rm A}^2\rho_{\rm A}c_{\rm s}(R_{\rm A})
\end{equation}
For M87 this implies an accretion rate
\begin{eqnarray}
\Mdot &\sim& 10^{26} [1 \kpc]^{\zeta} c^{4-2\zeta} \left[\frac{n_{\rm
e}(1 \kpc)}{0.24}\right] \\ \nonumber &&\left[\frac{c_{\rm
s}}{3.4\times10^7}\right]^{2\zeta -3}\left[\frac{R_{\rm S}}
{ 8.9\times 10^{14}}\right]^{2-\zeta} \gps \\ \nonumber 
& \sim &1 \Msunpyr. \nonumber
\end{eqnarray}
The temperature of the interstellar medium, $T$, is assumed to be that
determined from the ASCA analysis of the soft X-ray emission (Table 3,
Paper I) for the Virgo ellipticals.\footnote{The temperature profile
in the inner regions is the major uncertainty and may differ
significantly due to local cooling or heating processes.} The ASCA
spectra for the central cluster galaxies include a significant amount
of emission from the extended cluster gas; the measured temperatures
are therefore more likely to reflect the virial temperatures of the
host clusters.  Previous studies of M87 (e.g. Stewart et al. 1984),
however, have shown that most of the gravitating mass in the central
regions is associated with the galaxy itself rather than the Virgo
cluster as a whole. For this reason, we assume a temperature $kT \sim
1 \keV$ for the central cluster galaxies (and for the deprojection
analysis of the X-ray data in the innermost regions). This should
better approximate the virial temperature of the galaxy potential in
the inner regions ($T$ should scale approximately with the square of
the stellar velocity dispersion; the optical velocity dispersions for
M87 and NGC 1399 are similar to those of NGC 4472 and NGC 4649 - Van
der Marel 1991 - which have $kT \sim 1 \keV$; Paper I). The inferred
Bondi accretion rates for NGC 4696 and NGC 1399 are of the same
magnitude ($\sim 1\Msun$ yr$^{-1}$) as for M87, with slightly lower
values for the Virgo ellipticals.
Note that the gas densities at $R_{\rm A}$, and the corresponding
accretion rates, suggest that the systems are accreting at close to
their upper limits. If we balance the cooling time (assuming a power
law approximation for the cooling rate; e.g. McKee \& Cowie 1977) of
the hot gas at $R \approxgt R_{\rm A}$ with the local free-fall time,
then, from the Bondi theory, the maximum  accretion rate
is $\Mdot \approxlt 3 M_9 T^{1.6}_7 \Msunpyr$, which is consistent
with the assumption that gas will stay hot at $R \approxlt R_{\rm A}$ and
given our estimated accretion rates. 


For the black hole masses determined for these galaxies
(Table~\ref{t:bh}; Magorrian et al. 1998) accretion at the Bondi rate
with a radiative efficiency of $\eta=0.1$ (as assumed for standard
accretion) would yield a luminosity ($\equiv L_{\rm Bondi}$) exceeding
$10^{46}\ergps$ for central cluster galaxies (see Table~\ref{t:bh}).
Observationally, the nuclei of these giant ellipticals are 4-6 orders
of magnitude less active (see ratios of $L_{\rm Bondi}/L_{\rm Obs}$ in
Table~\ref{t:bh} and figures in Section 5). The observed luminosities
of their cores do not exceed $10^{42}\ergps$ for centre cluster
galaxies and a few $10^{40}$ for the Virgo ellipticals (see also DM99
and the IR limits from Paper I).  This provides strong observational
evidence for low radiative efficiency accretion.

\subsection{The spectrum of the accretion flow: testing the physics of
low radiative efficiency accretion}

In a low efficiency accretion flow around a supermassive black hole,
the majority of the observable emission is in the radio and X--ray
bands. In the radio band the emission results from cyclo--synchrotron
emission due to the near equipartition magnetic field in the inner
parts of the accretion flow. The X-ray emission is due either to
bremsstrahlung or inverse Compton scattering. The cleanest test for
probing the structure of low radiative efficiency accretion flows
(i.e., ADAFs) is to examine the correlation between the radio and X-ray
emission. 

The self--absorbed synchrotron spectrum in an ADAF slowly rises with
frequency in the radio band, up to some critical turnover frequency,
typically in high radio to sub--mm frequencies, above which the
emission should drop off quickly.  The peak emission always arises
from close to the black hole and reflects the properties of the
accreting gas within a few Schwarzschild radii.  Its Comptonization,
which can dominate the X-ray emission, is also produced in the very
inner regions of the flow.  Bremsstrahlung emission, on the other
hand, is typically produced at all radii in the flow.

The spectrum of a low efficiency accretion flow with a strong wind can
be characterized by a variable accretion rate such that $\Mdot \propto
r^{p}$ where $p\sim 1$ (i.e. a large fraction of the mass supplied is
carried away by the wind at large distances and only a small fraction
is accreted). This implies a strong suppression of the synchrotron and
Comptonized emission. However, relatively strong X-ray emission via
bremsstrahlung can still occur (e.g. DM99; Quataert \& Narayan 1999,
hereafter QN99).

Assuming that the wind is spectrally unimportant (i.e. non-radiating)
the importance of mass loss can be readily assessed by analyzing the
correlation between the radio and X-ray emission in elliptical
galaxies. If no wind is present and the accretion rate is of order the
Bondi rate, the spectra should exhibit prominent synchrotron emission
in the radio band and be dominated by Comptonization of this component
in the X-ray band. Such systems would have relatively soft X-ray
spectra.  If, on the other hand, a significant outflow is present, the
X-ray luminosity should dominate the radio luminosity and the X-ray
spectrum should be very hard, as expected from bremsstrahlung
emission.

Previous work has emphasized the power of radio and sub-mm
observations of nearby ellipticals to test ADAF models.  Such
observations allow both the synchrotron flux and the position of the
peak to be measured (DM99). In such work it was found that the
predicted radio emission, based on the ADAF model (for magnetic fields
in equipartition with thermal pressure and accretion at the Bondi
rate), exceeds the measured fluxes by 2-3 orders of magnitude.  Models
with strong mass loss (winds), which can easily accommodate the
observed suppression of the synchrotron emission, still predict a
significant X-ray flux due to bremsstrahlung emission from the outer
regions of the flows (if the accretion rates in these systems are
similar to the Bondi value).  The lack of any previous detection of
nuclear X-ray emission (DM99), however, left open the possibility that
the accretion rates in these systems were simply much lower than the
Bondi value.

The discovery of hard power-law X-ray emission (Paper I) in the Virgo
ellipticals studied by DM99 at the level of $10^{40} \ergps$, and in
three more central cluster galaxies with luminosities of up to
$10^{42} \ergps$, provides us with strong motivation to further
consider low-radiative efficiency accretion models and in particular
the evidence for mass outflow (note that throughout this paper we
assume that the majority of the observed X-ray emission is from
accretion onto the central supermassive black holes in these objects;
see Section 4 of Paper I for the justification of this
interpretation).

\section{The spectrum of the core emission}

In order to examine the nature of accretion in the present sample of
elliptical galaxies, we have compiled the best observational limits on
the broad band spectra of their core emission.  Our aim is to obtain
good observational limits on the core flux over a wide range of
frequencies rather than to compile a comprehensive list of all
previous observations. Some contribution from the jets (particularly
for the central cluster galaxies) and from the underlying galaxy are
unavoidable; in some sense, then, the derived spectra should be
considered as upper limits to the emission from the accretion flow.

The most up to date spectral energy distributions (SEDs) for the three
Virgo elliptical galaxies (NGC 4649, NGC 4636 and NGC 4472) were
compiled by DM99 (e.g see Tables 4,5 and 6 in DM99).  These include
stringent limits obtained at high radio and sub-millimeter frequencies
with the Very Large Array (VLA) and SCUBA, respectively. The X-ray
observations were upper limits obtained from ROSAT HRI.  Here they are
replaced with the ASCA detected hard power-law X-ray emission
described in Paper I (the spectral slopes and luminosities in the 1-10
\keV band are listed in Table 4 of that paper).

The spectral energy distribution for M87 (NGC4486) was previously
compiled by Reynolds et al. (1996). Again, the X-ray observations have
been updated to include the new ASCA (1-10 \keV) data (Paper I, Table
4).  We summarize the SEDs for NGC 1399 and NGC 4696 in
Tables~\ref{t:n1399} and ~\ref{t:n4696} respectively.  The data for
NGC 1399 also include new high radio frequency VLA data (consistent with a
significant suppression of the high energy emission with respect to
the standard ADAF model; Di Matteo, Carilli \& Fabian in preparation).
NGC 4696, the most luminous center cluster galaxy, is not as well
constrained in the radio band and lacks a black hole mass measurement.
According to the relationship $M_{bh} \propto M_{bulge}$ (e.g.,
Kormendy \& Richstone 1995; Magorrian et al 1998; Ho 1998), however,
it is expected to possess the the most massive black hole in our
sample.  Indeed, ASCA observations of NGC 4696 (Paper I) identify it
as having the most luminous hard X-ray power-law emission.

\subsection{Models}

Accretion from the hot interstellar medium in elliptical galaxies
(which should have relatively low angular momentum) may proceed
directly into a hot, advection-dominated, regime.

When gas is supplied to a black hole at rates well below the Eddington
rate it may not be able to radiate efficiently.  In an advection
dominated flow, which occurs at accretion rates $\Mdot \approxlt
\alpha^2 \Mdot_{\rm Edd}$, viscous stresses are assumed to dissipate
most of the energy locally into the ions (a number of investigations
have been carried out to study magnetic dissipation and particle
heating in plasmas appropriate to ADAFs; e.g.  Bisnovatyi-Kogan \&
Lovelace 1997; Quataert 1998; Gruzinov 1998; Blackman 1999; Quataert
\& Gruzinov 1999). A small fraction of the ion thermal energy is
transferred to the electrons (which are responsible for the radiation)
by Coulomb collisions, at a rate $\propto \Mdot/\Mdot_{\rm
  Edd}\alpha^{-2}$.  As a result of this poor thermal coupling (and
the assumed ion heating), the radiative efficiency of the flow can be
much less than the ``canonical'' disk value of $\sim 10\%$.

Narayan \& Yi (1994,1995a) noted that ADAFs have the interesting
property that the Bernoulli parameter, a measure of the sum of the
kinetic energy, gravitational potential energy, and enthalpy, is
positive over much of the flow; since, in the absence of viscosity,
the Bernoulli parameter is conserved on streamlines, the gas can, in
principle, escape to ``infinity'' with positive energy.  Narayan \& Yi
speculated that this might make ADAFs a natural candidate for
launching the outflows/jets seen to originate from a number of
accretion systems.  The positivity of the Bernoulli parameter, $Be > 0$
arises because viscous stresses transport energy and angular momentum
from small radii to large radii in the flow (cf Narayan \& Yi 1995 or
Blandford \& Begelman 1999).  Consequently, the positivity of $Be$,
and the resulting unboundedness of the ``accreting'' gas, is likely a
generic feature of any hot, low radiative efficiency, accretion model.

Blandford \& Begelman (1999) have recently suggested that mass loss
via winds in ADAFs may be both dynamically crucial and quite
substantial.  They constructed self-similar ADAF solutions in which
the mass accretion rate in the flow varies with radius $R$ as $\dot M
\propto R^p$ (see also simulations by Stone et al. in preparation). If
the wind carries away roughly the specific angular momentum and energy
appropriate to the radius from which it is launched, they show that
the remaining (accreting) gas has a negative Bernoulli parameter only
for large values of $p \sim 1$.  Essentially, Blandford \& Begelman
suggest that $Be > 0$ is unstable.  Matter will be lost to an
outflow/wind in sufficient quantity to guarantee that the remaining
matter is in fact bound, i.e., has $Be < 0$.  In their view, then,
outflows are inevitably driven by an accretion flow which cannot
radiate.  The outflow is necessary to guarantee that the accreting
matter is bound to the central object.  As we discuss in this paper,
the radio observations of DM99 and the X-ray observations of Paper I
provide support for this conjecture.

Numerical modeling techniques for ADAFs have advanced significantly
during the last two years (see Narayan, Barret \& McClintock 1997 and
Esin McClintock \& Narayan 1998). The calculations presented in this
paper use such models and also include some modifications. One
relevant difference here is the assumption that the mass inflow rate
satisfies
\begin{equation} 
\Mdot = \Mdot_{out}\left(\frac{R}{R_{\rm out}}\right)^p  \ \ \ {\rm for} \ \ \  R < R_{\rm out},
\end{equation}
as in previous models by DM99 and QN99 ($\dot M = \dot M_{\rm out}$
for $R > R_{\rm out}$). The quantity $\Mdot_{\rm out}$ is the
accretion rate at the radius $R_{\rm out}$ where winds become
important.  Note that $R_{\rm out}$ need not be the same as $R_A$ (the
accretion radius), the radius at which the accretion flow starts.

The most relevant difference in the code is an improved calculation of
the bremsstrahlung emissivity. Following previous work (Di Matteo \&
Fabian 1997; DM99) we now calculate the spectral emissivity using the
expressions for the electron-electron bremsstrahlung and electron-ion
bremsstrahlung computed by Stepney and Guilbert (1983). For the
evaluation of the electron-ion bremsstrahlung (the dominant
contribution), it is important to use the cross section in the Born
approximation for a relativistic electron in a Coulomb field given by
Gould (1980). For electron-electron bremsstrahlung process, the
cross-section is much more complicated and we have used the
numerically integrated expression given in Stepney and Guilbert
(1983). The resulting emissivity includes both contributions.

For convenience and consistency with previous work we scale black hole
masses in solar units, $M = m\Msun$, and accretion rates in Eddington
units, $\Mdot=\mdot \Mdot_{\rm Edd}$ and $\Mdot_{\rm Edd}=L_{\rm
  edd}/0.1c^2$.  We rescale the radial co-ordinate in Schwarzschild
units, $R=rR_{\rm S}$.

The ADAF model is specified by four parameters: $\alpha$, the
viscosity parameter, $\beta$, the ratio of gas to magnetic pressure,
$\gamma$, the adiabatic index of the fluid, which is assumed to be
$5/3$, and $\delta$, the fraction of the turbulent energy which heats
the electrons.  The thermodynamic state of the flow is described by
the ion temperature $T_{\rm i} \sim 10^{12} \beta r^{-1}\K$ (which is
to a very good approximation the virial temperature), the electron
temperature, $T_{\rm e}$, and the magnetic pressure.  The electron
temperature profile is obtained by solving the full electron energy
equation, including the electron entropy gradient (Nakamura et al.
1997; Esin et al. 1997; Narayan et al. 1998)

The microphysical parameters in ADAFs are quite uncertain; unless
otherwise specified we take $\beta=10$, $\alpha=0.1$ (in accordance
with the viscosity parameter scaling roughly as $\alpha \sim 1/\beta$,
as expected if turbulent stresses arise solely from magnetic fields;
Hawley, Gammie \& Balbus 1996), and $\delta = 10^{-2}-0.1$ (as we
shall show later the observational constraints allow us to vary the
fraction of electron heating with no significant modification to any
of the conclusions). Excepting $\alpha$ (which is discussed in \S 7),
the particular values assumed for these parameters are not important
for any of our qualitative conclusions.  Variations in the predicted
spectrum due to changes in the microphysical parameters have been
investigated by QN99.

The other model parameters are $m$, which is estimated from
observations (apart from NGC 4696), $p$, $r_{\rm out}$ and $\mdot_{\rm
out}$.  $\mdot_{\rm out}$ is adjusted to reproduce the X-ray data.
The predicted spectrum depends primarily on the fraction of the
incoming mass accreted onto the central object, so that $p$ and
$r_{\rm out}$ are somewhat degenerate.  They can, however, be
decoupled to some extent when one attempts to satisfy both the radio
and X-ray constraints.

\section{Spectra of ADAFs with winds}

ADAF models with a variable accretion rate have been investigated by
DM99 in the context of elliptical galaxy nuclei and in greater detail
and generality by QN99. Here we will briefly summarize how the
different spectral features are affected by the presence of outflows.
In the magnetized, optically thin plasma of an ADAF, with an electron
temperature of $10^9-10^{10} \K$, the most important radiation
processes are synchrotron emission, Compton scattering and
bremsstrahlung emission.

We will use the self-similar approximation to indicate how the
introduction of a radially varying $\mdot$ (see Eqn.~1) changes the
predicted spectrum. As mentioned earlier, in the thermal plasma of an
ADAF, synchrotron emission rises steeply with increasing frequency.
Under most circumstances the emission becomes self--absorbed and gives
rise to a black--body spectrum below a critical frequency, $\nu_{\rm
c}$. Above this frequency it decays exponentially as expected from a
thermal plasma.  The peak frequency scales as $\nu_{\rm c} \propto B
T_{\rm e}^2 \propto m^{-1/2}\mdot^{1/2} T_{\rm e}^2 r^{-5/4}$ and the
luminosity (approximated by the Rayleigh--Jeans limit) varies as
(Mahadevan 1997)

\begin{equation}
\nu_{\rm c} L_{\nu c} \propto \nu_{\rm c}^3T_{\rm e}m^2r^2 \propto
\beta^{-3/2}T_{\rm e}^7\mdot^{3/2}. \label{lsynch}
\end{equation} 
Note that emission observed at higher frequencies originates at
smaller radii and that the total power is a very strong function of
temperature.

For all other model parameters fixed, the predicted synchrotron
emission decreases strongly with the introduction of a wind (i.e.,
with increasing $p$).  There are two reasons for this.  First,
increasing $p$ decreases the magnetic field strength near $r \sim 1$,
where the high frequency synchrotron emission originates (this is
because the density and the gas pressure decrease as $p$ increases).
In addition, the electron temperature decreases as $p$ increases.
This is because adiabatic compression of the electrons is less
efficient (when $p$ is large, the density profile is flatter, and
hence $T_e$ is smaller). By equation (\ref{lsynch}), the synchrotron
emission is particularly sensitive to the electron temperature.
Therefore, the synchrotron emission falls very rapidly with increasing
$p$.

The Compton power, i.e., the Compton scattering of soft synchrotron
photons by the hot electrons in the accreting gas, decreases with
increasing $p$ even more strongly than the synchrotron does.  The
optical depth to electron scattering decreases with the introduction
of a wind, as does the electron temperature. In other words, the
Compton-$y$ parameter $ \sim 16(kT_{\rm e}/m_{\rm e}c^2) \tau$ is $\ll
1$ in ADAFs with outflows.

In~contrast~to~the~processes discussed above, bremsstrahlung emission
arises from all radii in the flow. The emission at frequency $\nu$
is, however, dominated by the largest radius which satisfies $h \nu
\sim k T(r)$.  The large radii dominate because, so long as $h \nu
\approxlt k T$, the bremsstrahlung luminosity from a spherical shell
of radius $r$ and thickness $dr \sim r$ is $\propto r^3 \rho^2
T^{-1/2} \propto r^{2p} T^{-1/2}$ which is an increasing function of
$r$ (even for $p = 0$).  Note that the radial dependence of the
emissivity depends on the density profile of the flow and therefore on
the strength of the wind.  This is seen explicitly in Figure 1 which
shows the bremsstrahlung emissivity, $\epsilon_\nu$ (ergs s$^{-1}$
Hz$^{-1}$), as a function of $r$ for 3 different X-ray energies (from
top to bottom, $1,10,100$ keV).  The
solid lines show a no-wind ADAF model while the dotted lines show an
ADAF model with $p = 0.5$ and $r_{\rm out} = 10^3$ (this model is
favored by the observations discussed in \S5).  Note that the soft
X-ray emission is always dominated by radii $\sim 10^3 - 10^4 R_S$ since this
is where $k T \sim $ a few keV.  Even in the absence of a wind, the
hard X-ray emission is dominated by radii $\approxgt 100 R_S$ since
$kT \approxgt 100$ keV for $r \approxlt 10^2-10^3$.  In the presence
of a wind, radii $\approxlt \ro$ contribute negligible to the
bremsstrahlung emission.

\begin{figure}
\centerline{
\psfig{figure=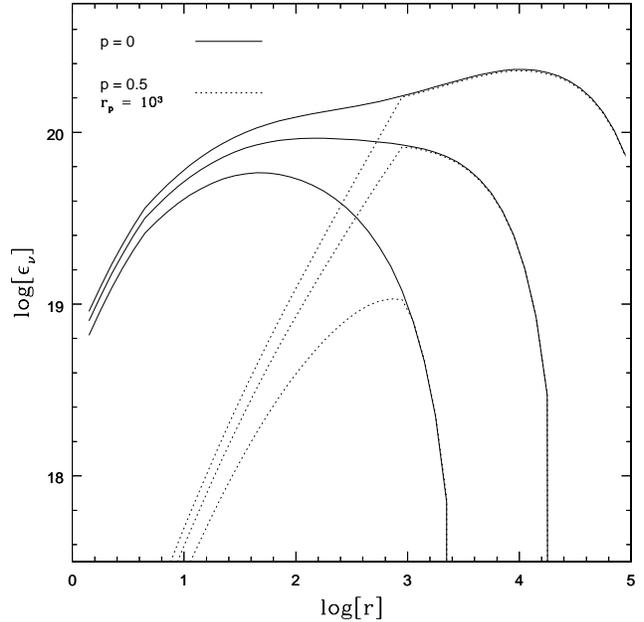,width=0.5\textwidth}
}
\caption{The bremsstrahlung emissivity, $\epsilon_\nu$ (ergs s$^{-1}$
  Hz$^{-1}$), as a function of $r$ for three different X-ray energies
  (from top to bottom, $1, 10,\ \& 100$ keV).  The solid lines
  show the bremsstrahlung emission for a no-wind ADAF model while
  dotted ones show an ADAF model with a wind characterized by $p = 0.5$
  and $r_{\rm out} = 10^3$ (observationally favored parameters; see
  Section 5).}
\label{bremfig}
\end{figure}

As pointed out by QN99 (and explicitly shown in Figure~\ref{bremfig}),
bremsstrahlung emission at $1- 10 \keV$ is rather insensitive to the
presence of a wind since it originates from the outer regions of the
flow. X-ray observations in this band can therefore directly measure
the density at the outer edge of the flow and provide a firm estimate
of the rate at which matter is fed to the accretion flow (i.e.
$\mdot_{\rm out}$).  At energies around $10 \keV$ and above, the
bremsstrahlung emission would decrease with increasing $p$ if the
outflow were to start at $r_{\rm out} \sim 10^4$.  The absence of a
decrease in the thermal bremsstrahlung emission at and above $10\keV$
(cf \S5) therefore implies either that there is no wind or that the
outflow starts at radii $\ll 10^4$.

\section{Comparison with the data}

Figure~\ref{spec} shows the observational constraints on the spectra of
the six giant elliptical galaxies together with our model predictions.
A striking feature of the broad band spectral energy distributions is
the ``non-thermal'' character of the emission. In particular, it is
apparent that most of the luminosity is produced in the X-ray band
(very hard energy spectral slopes are observed ranging from
$\alpha_{x} \sim 0.4$ to $-0.3$), with roughly two orders of magnitude
less luminosity coming out in the radio band. (The case of M87 is less
drastic, with the X-ray luminosity exceeding that in the radio band by
less than a factor 10. The results for NGC 4696 are also more
uncertain due to the lack of any high frequency radio data). The SEDs
are very different from those of standard AGN, which are typically
much flatter (in $\nu L_{\nu}$) with big blue bumps in the optical/UV
bands (generally interpreted as thermal emission from a standard thin
disk) and X-ray emission with slopes of $\alpha_{x} \sim 0.8-1$.

In the following Sections, we present detailed models for each of the
galaxies in our sample. The results shown in the figures, for various
values of $p$, all have $\alpha = 0.1$, $\beta =10$ and $\delta =0.01$
(or 0.1).  The value of $\mdot_{\rm out}$ is adjusted to fit the
observed ASCA X-ray flux at $1$ keV.  These measurements are then
compared to the predicted Bondi accretion rates to check for
consistency. In all of the models $p$ is required to be large enough
to satisfy the stringent radio limits (for NGC 1399, NGC 4649, NGC
4472 and NGC 4636).  We shall show that this also yields a
bremsstrahlung dominated X-ray spectrum (as required independently by
the ASCA data).  We start by describing the spectral models for each
object and then discuss the general picture which arises from the
comparison with the data.

\subsection{Central cluster galaxies}
\subsubsection{M87}
Figure~\ref{spec}a shows the spectral models for M87 for different
values of $p$ and $r_{\rm out}$.  The high frequency radio data
(unlike in most of the other galaxies) is completely consistent with a
model with no wind, i.e. $p=0$. In this case, however, the X-ray
emission is due to Comptonization of the synchrotron photons.  The
predicted slope is too soft to be consistent with the ASCA data
(dashed line Fig.~\ref{spec}a). The accretion rate in this case is
$\mdot=\mdot_{\rm out} \sim 10^{-3}$.  This model also predicts an
optical flux (due to the first Compton bump) which exceeds the HST
limit: a further increase in $\mdot$, which seems to be required by
the X-ray data, would significantly violate this limit.

The dotted line shows a model for $r_{\rm out} =10^4$ and $p = 0.37$
(which has $\dot m_{\rm out} = 0.025$).  Although this model has the
correct hard slope up to $\sim 5-7 \keV$, the effects of a wind
starting at $r = 10^4$ introduce a fairly strong suppression of the
X-ray emission above $\sim 5 \keV$ (a spectral break in the
bremsstrahlung emission as described in \S4). The best fit model
(solid line) shows the effects of reducing $r_{\rm out} $ to $\sim
300$ (with $p = 0.45$ and $\dot m_{\rm out} = 0.015$).  In fact, good
agreement with the X-ray data is obtained for any $r_{\rm out}
\approxlt 1000$.  The quoted values for $r_{\rm out}$ should only be
considered accurate to order of magnitude, but values of $\sim
10^2-10^3$ are preferred for all systems we consider (see the
comparison between the solid, dotted and dashed lines in
Fig.~\ref{spec}a).

In the models with winds, the 100 GHz VLBI data might require an
additional component; the synchrotron jet or the wind itself could be
candidates; note also that increasing $\delta$ (with
a relative increase in $p$ so that the X-ray emission is still
dominated by bremsstrahlung), \eg, further electron heating in
reconnection sites, would increase the electron temperature and
therefore the synchrotron flux by a factor $\propto T_{\rm e}^7$.
This might produce the required radio flux as well (see also
Fig.\ref{spec}d for NGC4649).

\subsubsection{NGC 1399}
Fig.~\ref{spec}b shows three spectral models for NGC 1399. Both the
high frequency VLA data and the ASCA spectral slope are inconsistent
with with the $p=0$ model (dashed line).  Also, the very hard X-ray
slope is best accounted for by a model with larger $p$ and smaller
$r_{\rm out}$ so that no spectral break is introduced around $\sim 6-
8 \keV$ (thinner solid line: $r_{\rm out} = 10^4$, $p=0.4$,
$\mdot_{\rm out} \sim 9 \times 10^{-3}$).  A model which is consistent
with both the suppression of synchrotron radiation in the radio and
the hard slope in the X-ray band has $r_{\rm out} =300$, $p = 0.6$,
and $\mdot_{\rm out} \sim 5 \times 10^{-3}$ (thicker solid line in
Fig.~\ref{spec}b).

\subsubsection{NGC 4696} 
We do not have an estimate of the black hole mass in NGC 4696. Given
that this is the most luminous and massive galaxy in the Centaurus
cluster we will assume that its black hole mass is comparable to that
of M87 or NGC 1399. In detail, we assume $m \sim 5 \times 10^9$ (which
is roughly consistent with the radio data).  No high frequency radio
data are available for this object and the only important constraint
is given by the ASCA observations. The relatively high flux and hard
X-ray ray slope can be accounted for by models with $p \sim 0.6$ and
$r_{\rm out} \sim 300$ (thicker solid line Fig.~\ref{spec}c).  This
implies $ \mdot_{\rm out} \sim 0.03$. Models with no wind would again
predict Compton dominated spectra with slopes too soft to explain the
ASCA observations (dashed line, Fig.~\ref{spec}c).

\subsection{Virgo ellipticals}

\subsubsection{NGC 4649}

NGC 4649 has the best observational constraints of any object in the
sample. The radio data for this object show a very point-like source,
dominated by its core component at high radio frequencies. The VLA 22
to 43 GHz data points and the SCUBA sub-mm upper limit at ($10^{11}
$Hz) also imply a very sharp spectral turnover in the radio spectrum
which is hard to explain with any non-thermal particle distribution,
but easily obtained from thermal synchrotron radiation (DM99). The
flux in the radio-sub-mm band is orders of magnitude below that
expected from models without winds, but is easily matched by models
with winds (see Fig.~\ref{spec}d and 5 in DM99). The value of $p$
required to account for the (lack of) radio emission ($\sim 0.4$)
would produce a significant spectral break in the high energy X-ray
band due to the suppression of the higher temperature (inner radii)
bremsstrahlung if $r_{\rm out} \sim 10^4$. Better agreement with the
data is again obtained by taking $r_{\rm out} \sim 300$ and $p\sim
0.5$, which imply $\mdot_{\rm out} \sim 2.7 \times 10^{-3}$ (thicker
solid line, Fig.~\ref{spec}d). The dash-dotted line shows a models for
$\delta = 0.1$.  The increased electron heating in models with winds
implies a slight increase in $p$ in order to re-establish agreement
with the radio constraints (p =0.8; a slight decrease of $r_{\rm out}$
might also be necessary for the high energy X-ray data). Note that
models with winds (as required by the observational constraints) can
easily allow for different fractions of electron heating with no
significant change to any of our conclusions (see QN99 for a general
discussion of this point).
  
\subsubsection{NGC 4472}
Although the VLA and SCUBA sub-mm data do not imply a very sharp
spectral turnover in this object, they require significant suppression
of the emission from the inner regions of the flow, as expected in the
presence of a strong wind.  These limits, together with the ASCA X-ray
flux and spectral index, also prefer models for which $r_{\rm out}
\sim 300$ and imply $p\sim 0.5$ and $\mdot_{\rm out} \sim 5\times
10^{-3}$ (see Fig.~\ref{spec}e).

\subsubsection{NGC 4636}
NGC 4636 is the only object in this sample with $m < 10^9$.  The
relatively small black hole mass, as estimated by Magorrian et al.
(1998), and the relatively high accretion rates required to fit the
ASCA X-ray flux measurement, shift the synchrotron peak to higher
frequencies than those expected by the radio and sub-mm measurements
(Fig.~\ref{spec}f dashed lines).\footnote{In DM99 the much lower
  accretion rate and differences in the electron temperature could
  accommodate the peak position as set by the data.}  In this case,
the X-ray spectral slope and flux do not constrain $r_{\rm out}$.
Models with $r_{\rm out} = 10^4$ ($p = 0.55$ and $\dot m_{\rm out} =
0.03$) and $r_{\rm out} = 300$ ($p = 0.9$ and $\dot m_{\rm out} =
0.018$) provide satisfactory fits to the X-ray data.

Note that an increase in the black hole mass to $m \sim 10^9$ (similar
to the other objects) would shift the synchrotron peak to lower
frequencies and into good agreement with the radio limits (solid lines
in Fig.~\ref{spec}f).  The very sharp spectral turnover implied by the
VLA observations can otherwise only be explained by a very peculiar
distribution of non-thermal particles with a very sharp cut off at
relatively low gamma or by an additional thermal synchrotron component
in the jet or wind.

We now turn to the general conclusions which can be drawn from our
modeling of the six nearby elliptical galaxies and a brief discussion
of the crucial properties of their environments and their influence on
the accretion flows.

\begin{table*}
\caption{Summary of data for the core of NGC 1399.}
\label{t:n1399}
\begin{center}
\begin{tabular}{ccccc}\hline
Frequency & $\nu F_{\nu}$ & reference & notes \\
$\nu$ (Hz) & (10$^{-15}$\,erg\,s$^{-1}$\,cm$^{-2}$) & \vspace{0.3cm} \\
\hline 
$2.3\times 10^9$ & 0.35 & Slee et al. (1994) & PTI \\
$4.9\times 10^9$ & 0.50 & Sadler et al. (1989) & VLA\\
$8.4\times 10^9$  & 1.68 & Slee et al. (1994) & PTI\\
$8.4\times 10^9$ & 1.81 & Di Matteo et al. (1999) & VLA\\
$2.2\times 10^{10}$ & 5.60 & Di Matteo et al. (1999) & VLA \\
$4.3\times 10^{10}$ & 7.10 & Di Matteo et al. (1999) & VLA \\
$2.4\times 10^{17}$  & 98  & Paper I & ASCA\\
$2.4\times 10^{18}$  & 1350 & Paper I & ASCA \\
\hline
 
\end{tabular}
\end{center}
\parbox {3.5in}
{PTI: Parkes-Tidbinbilla interferometer}
\end{table*}

\begin{table*}
\caption{Summary of data for the core of NGC 4696.}
\label{t:n4696}
\begin{center}
\begin{tabular}{ccccc}\hline
Frequency & $\nu F_{\nu}$ & reference & notes \\
$\nu$ (Hz) & (10$^{-15}$\,erg\,s$^{-1}$\,cm$^{-2}$) & \vspace{0.3cm} \\
\hline 
$2.3\times 10^9$ & 0.57 & Slee et al. (1994) & PTI \\
$4.9\times 10^9$ & 2.70 & Sadler  et al. (1989) & VLA\\
$8.4\times 10^9$  & 1.17 & Slee et al. (1994) & PTI\\
$2.4\times 10^{17}$  & 358  & Paper I & ASCA\\
$2.4\times 10^{18}$  & 6221 & Paper I & ASCA \\
\hline
 
\end{tabular}
\end{center}
\parbox {3.5in}
{PTI: Parkes-Tidbinbilla interferometer}
\end{table*}

\section{Low radiative-efficiency accretion and cooling flows in ellipticals}

The broad band spectra of the six nearby elliptical galaxies are
characterized by one or both of the following properties: a large
X-ray to radio luminosity ratio (with indications of a sharp spectral
cut-off in the radio) and a hard X-ray spectral index.  These point to
the importance of outflows in low luminosity systems.  The strong
synchrotron peak in the radio, expected to dominate the broad band
emission, is strongly suppressed.  The X-ray emission is typically
extremely hard and is well explained by bremsstrahlung
emission.\footnote{The X-ray spectra of several of the sources are
actually slightly harder than expected from thermal bremsstrahlung
emission (i.e. in $\nu L_{\nu} \propto \nu$ for $kT_{\rm e} <
h\nu$). We expect that small amounts of intrinsic absorption are
responsible, since the observed slopes are too hard to arise from any
other likely process e.g. photon starved thermal Comptonization, or
non-thermal Comptonization; see also the infra-red (IRAS) limits
presented in Paper I.}  As discussed in Section 4, the introduction of
a wind accounts for these characteristic spectral features.  Note also
that these conclusions are not dependent on the choice of the
microphysical parameters (although high values $\alpha$ are required
for consistency); in particular it is worth noting that significant
electron heating ($\delta \approxgt 0.1$) is perfectly viable.

The only object which does not show a spectral turnover at high radio
frequencies and which therefore does not require a mechanism to
suppress the radio emission is M87 (apart from NGC 4696, for which no
data is available to constrain the high frequency emission). Recall
that this is the only object in our sample with a one sided jet which
is most probably relativistically beamed towards us (and, in
particular, M87 is among the more active FR-I radio sources in central
cluster galaxies) . As shown by the radio map of Baath et al. (1992)
the jet is resolved by high resolution VLBI 100 GHz observations
together with the core. As we shall discuss later, contributions from
such a jet might even be important for what we are labeling accretion
flow emission.  Nevertheless, models which do not include significant
mass loss are unable to explain the hard X-ray spectral slope, which
is easily accounted for by bremsstrahlung emission.

\subsection{Outer Radii and Angular Momentum}

The amount of mass loss required to satisfy the X-ray to radio flux
ratio and/or the observed X-ray slope would give rise to a cut--off in
the bremsstrahlung emission at energies $\approxlt 10 \ \keV$
(apparently inconsistent with the ASCA data) were the outflow to start
at radii $r \approxgt 10^4$. Better agreement with the X-ray
observations is obtained for outflows starting inside the outer
boundary of the flow at radii $\approxlt 10^3$.  

This result may suggest that angular momentum does not dominate the
flow close to the accretion radius, perhaps because angular momentum
in the hot gas is transported outward in the halo by turbulence
(Nulsen, Stewart \& Fabian 1984) or because the hole is moving with
respect to the gas. (In order that the gas shed at large radii may
flow unimpeded to the centre of the galaxy, its angular momentum must
be dissipated. A rotating, contracting flow - as a cooling flow in an
elliptical galaxy - is very likely to become turbulent. Angular
momentum must then be transported efficiently outside the medium and
is likely to be taken up by some parts of the shed gas -- as the gas
flows in).
The inflow across the accretion radius could then be roughly
spherically symmetric with the specific angular momentum being a
fraction, $\xi$, of the one required for a Keplerian orbit at $R
\approxlt R_{\rm A}$. In particular, in order for the angular momentum
dominated flow to start at $r \sim r_{ out} \sim \xi^{2} r_{A} \sim
10^3$, $\xi$ must be $\approxlt 0.03$ (radial accretion is precluded
unless $\xi \le 10^{-3}$). Further investigations both on the
theoretical and observational side are necessary to more carefully
assess the values of $r_{\rm out}$ and the above hypothesis.

We now briefly discuss how the gas densities (i.e the rates at which
matter is fed at $r_{\rm out}$) required to explain the observed X-ray
fluxes compare to the expected Bondi rates.

\begin{table*}
\caption{Model parameters for Fig.~\ref{spec}}
\label{t:para}
\begin{center}

\begin{tabular}{ccccc}\hline
Object & $r_{\rm out}$ & $p$ & $\mdot_{\rm out}$& Line type \\
\vspace{0.3cm} \\
\hline

M87  & $10^4$ & 0 & $10^{-3}$ & dashed \\
     & $10^4$ & 0.37 & $ 0.03$ & thin solid \\       
     & 800  & 0.4  & 0.017 & dotted  \\
     & 300  & 0.45  & 0.015 & thick solid\\
     & 100  & 0.5   & 0.013 & dotted \\  
NGC 1399 & $10^4$& 0 & $0.7\times 10^{-3}$ & dashed \\
         & $10^4$& 0.44 & 0.01 & thin solid \\
         & $300 $ & 0.6& $6.7 \times 10^{-3}$ & thick solid \\ 
NGC 4696 & $10^4$& 0 & $1.4 \times 10^{-3}$ & dashed \\
         & $10^4$& 0.4 & 0.05 & thin solid \\
         & $300$& 0.62 & 0.03 & thick solid \\
NGC 4649 & $10^4$ & 0 & $4 \times 10^{-4}$ & dashed \\
         & $10^4$ & 0.38 & $4.4 \times 10^{-3}$ & thin solid \\
         & 300 &   0.52 & $2\times 10^{-3}$ & thick solid \\
	 & 100 &   0.8$(\delta=0.1) $& $2.5\times 10^{-3}$ $(\delta=0.1)$& dot-dashed \\
NGC 4472 & $10^4$ & 0 & $5 \times 10^{-4}$ & dashed \\
         & $10^4$ & 0.37& $8\times 10^{-3}$ & thin solid  \\ 
         & 300 & 0.5 & $5\times 10^{-3}$ & thick solid \\
NGC 4636 & $10^4$ & 0 & $10^{-3}$ $(4 \times 10^{-4})$ & dashed\\
         & $10^4$ & 0.55 (0.5) & 0.03 ($6\times10^{-3}$) &thin solid (short dashed \\
         & 300 & 0.9 (0.62) & 0.018 ($3\times 10^{-3}$) &thick solid (short dashed\\ 
\hline

\end{tabular}
\end{center}
\parbox {7in} {Notes: The values of $p$ and $\mdot_{\rm out}$ in
brackets, for NGC 4636, refer to those obtained assuming $m \sim
10^9$, see also Fig~\ref{spec}f.}
\end{table*}

\subsection{Accretion Rates and magnetic fields in the central regions}

\begin{figure*}
  \centerline{ \hbox{ \vbox{ \vspace{-2.0cm}
        \psfig{figure=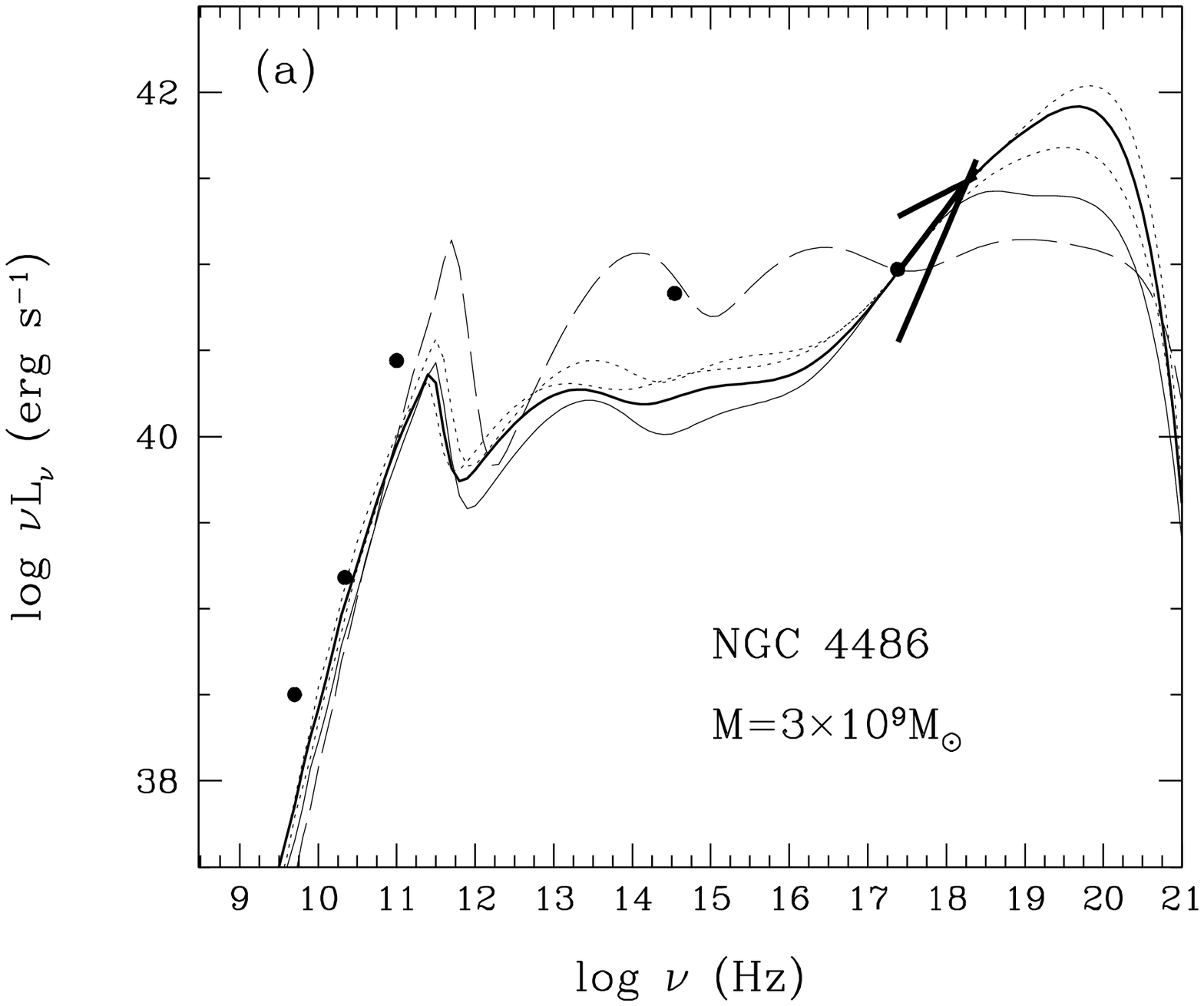,width=0.467\textwidth} \vspace{-1.6cm}
        \psfig{figure=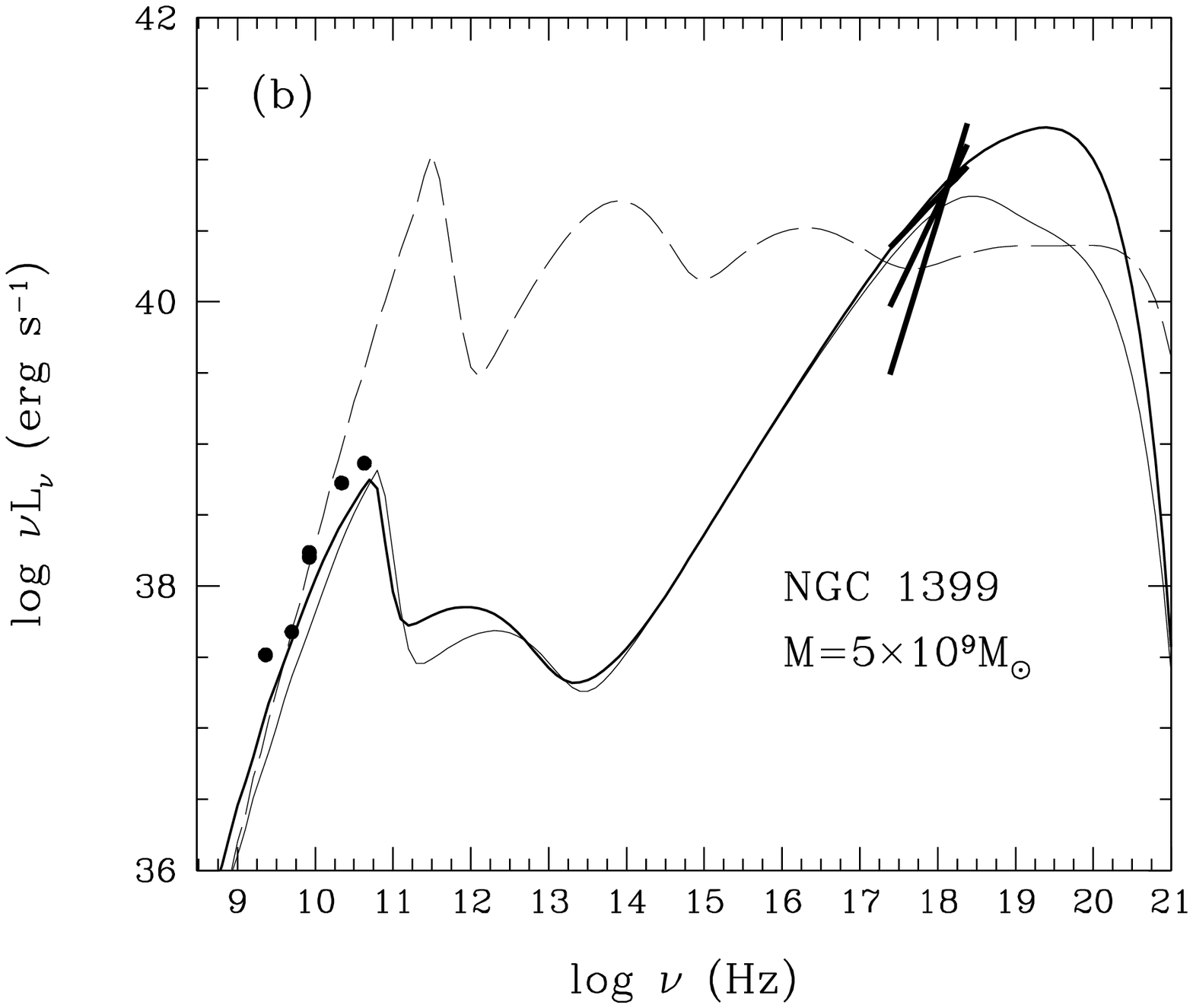,width=0.467\textwidth} \vspace{-1.6cm}
        \psfig{figure=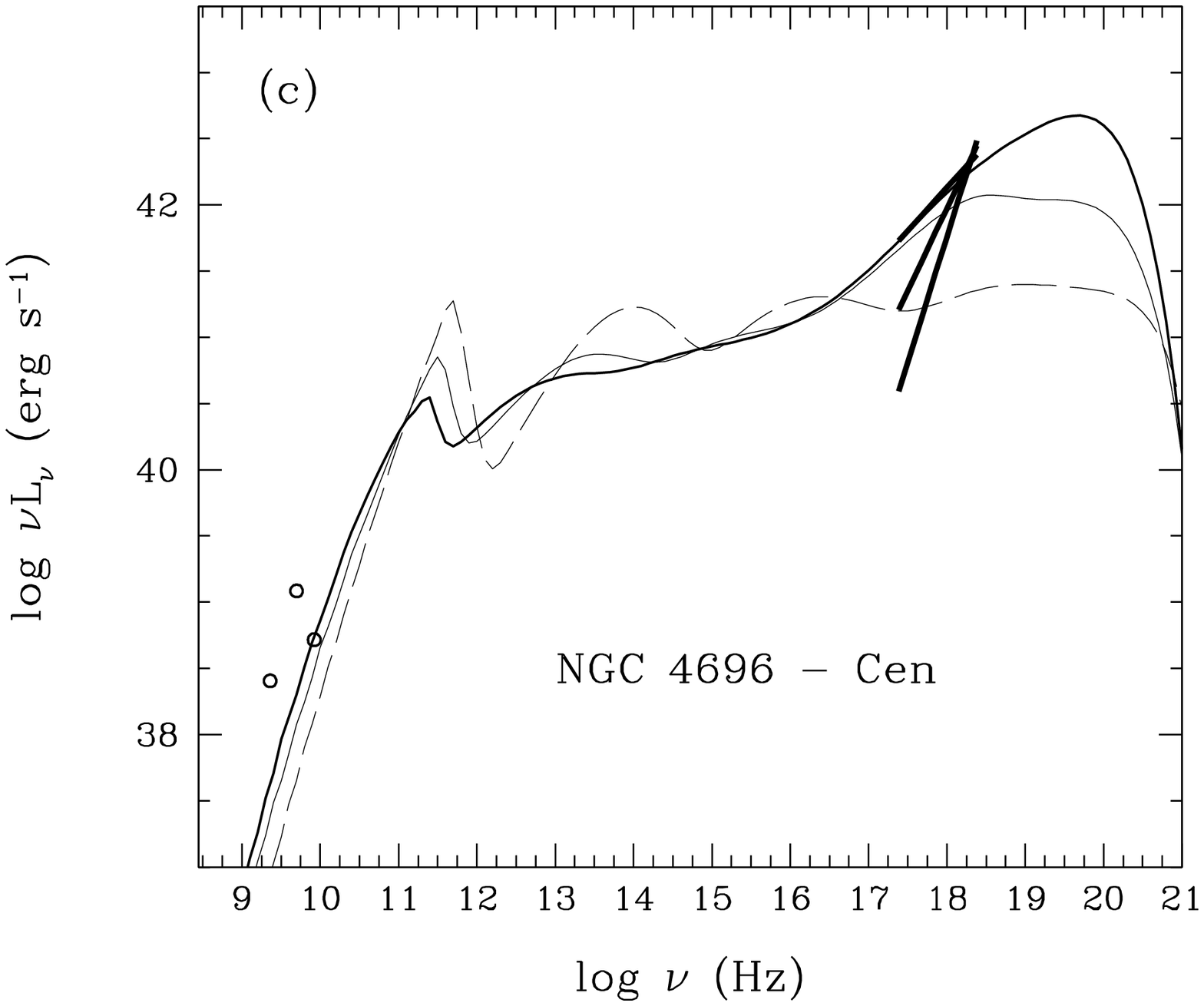,width=0.467\textwidth} } \vbox{
        \psfig{figure=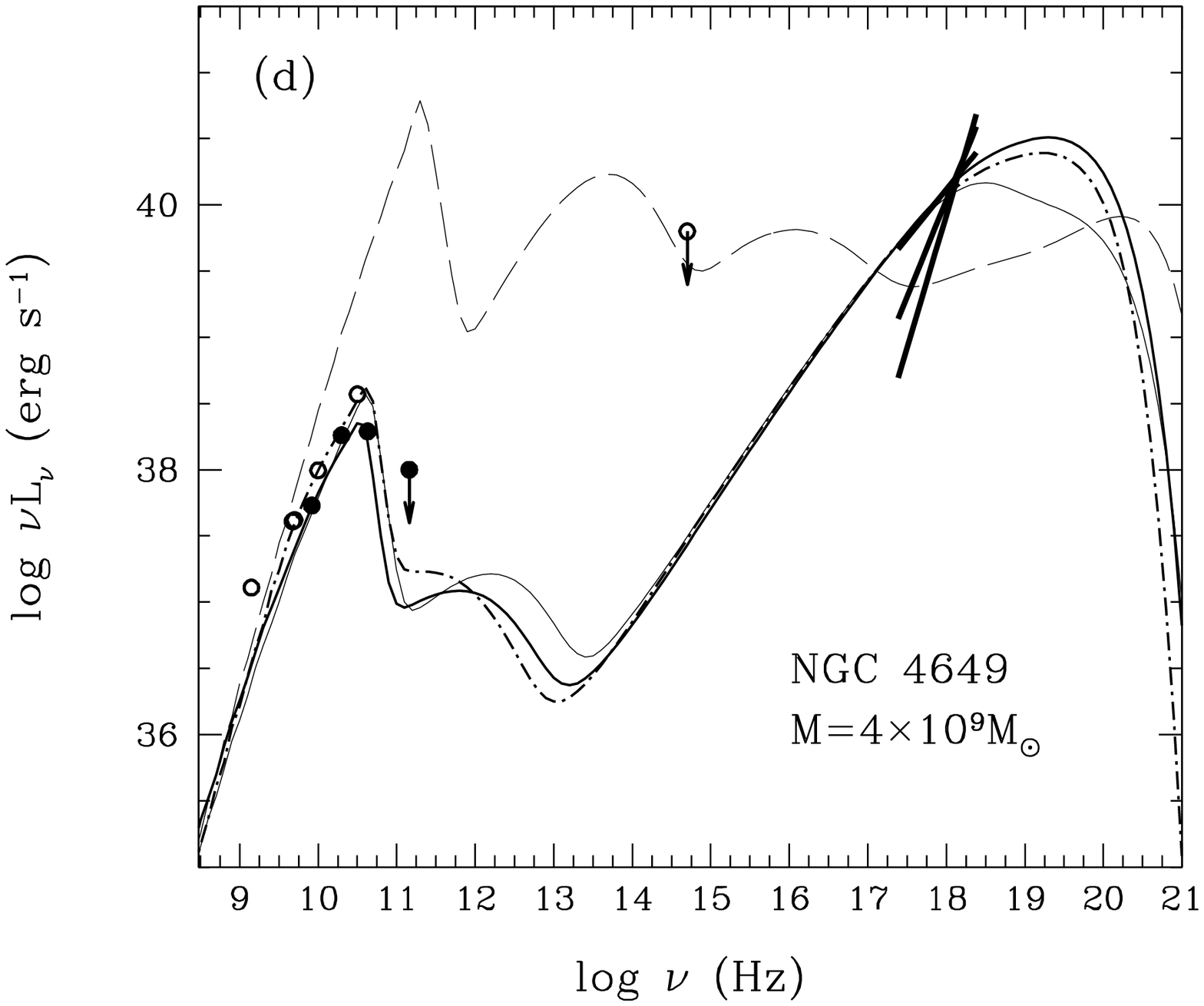,width=0.467\textwidth} \vspace{-1.6cm}
        \psfig{figure=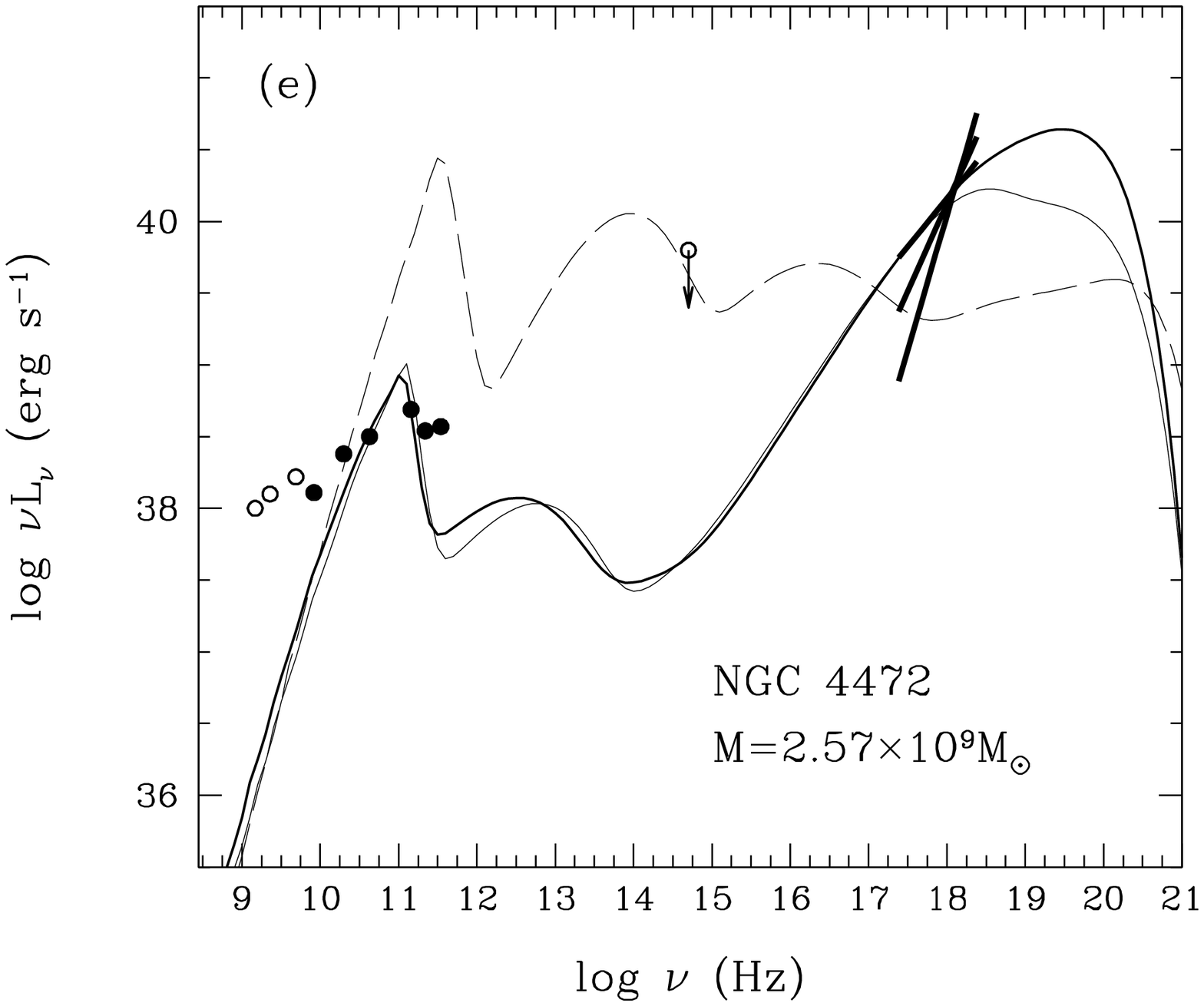,width=0.467\textwidth} \vspace{-1.6cm}
        \psfig{figure=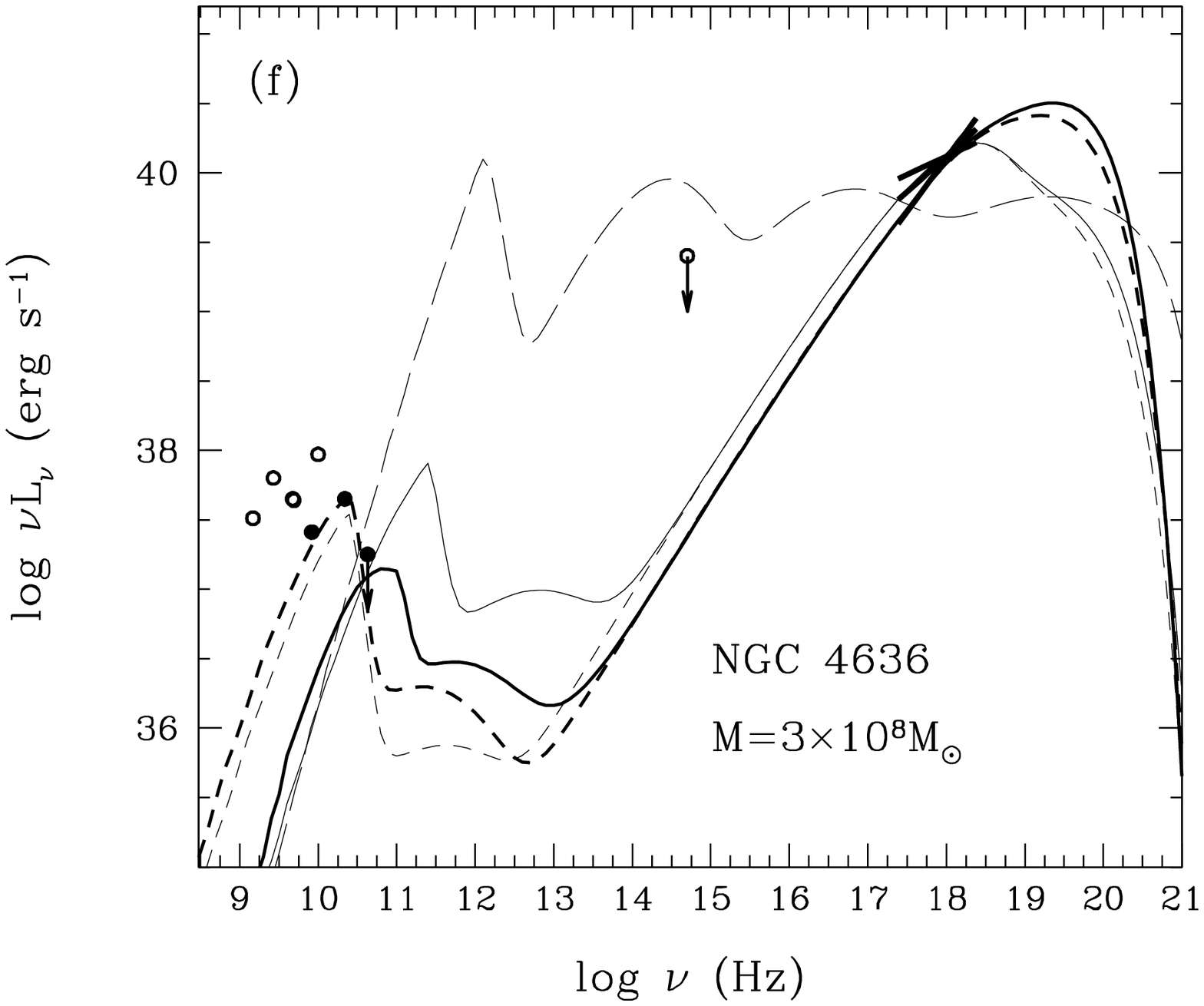,width=0.467\textwidth} }}}
  \vspace{-1.0cm}
\caption{Spectral models calculated for ADAF models with and without
winds. The central cluster galaxies are on the left-hand side and the
Virgo elliptical galaxies are on the right-hand side. The model
parameters, corresponding to the fits, are tabulated in
Table~\ref{t:para}.  The long dashed line represents no-wind ADAFs;$ p
=0$.  The thicker solid line, models for $r_{\rm out} \sim 300$.  The
thinner solid line, models for $r_{\rm out} \sim 10^4$.  The dotted
lines in panel above and below the solid line (a) are for $r_{\rm out}
\sim 100$ and $r_{\rm out} \sim 800$ respectively.  The short dashed
lines in panel (f) assume $m\sim 10^9$. The dash-dotted line in (d) is
for a model with $\delta =0.1$. Such models show better agreement
with the characteristics radio spectrum.  The solid dots represent the
best constraint on the core emission.  The thick solid lines the
slopes and fluxes measured form the ASCA analysis (paper I), For the
Virgo Cluster galaxies we assume a fixed luminosity distance of
18Mpc. For NGC 1399 and 4696 the luminosity distances are calculated
using the redshifts and an assumed cosmology of $H_0$=50 \kmpspMpc,
$\Omega = 1$ and $\Lambda = 0$ (29 and 63 Mpc respectively.}
\label{spec}
\end{figure*}

The accretion rates for each model are obtained by normalizing the
model to the measured X-ray fluxes. The results for all of the models
shown in Fig.~\ref{spec} are tabulated in Table~\ref{t:para}.  The
accretion rates required to explain the X-ray fluxes are consistent
with the Bondi values predicted in Section 1.1 from the deprojection
analysis of the X-ray gas. The differences in accretion rates, for a
given object and X-ray flux, with varying $r_{\rm out}$ and $p$ are
mostly from the slightly different electron temperature profiles in
such models.  We also note that the accretion rates implied by models
with no wind would be smaller by roughly an order of magnitude.  The
X-ray flux in all such models is, however, due to inverse Compton
scattering of the synchrotron photons.  In no case can a
Compton-dominated ADAF model give a satisfactory fit to the radio and
X-ray observations.

The Bondi accretion rates previously estimated for M87 by Reynolds et
al. (1996) and for the Virgo ellipticals by DM99 are typically a
factor 5-8 lower than the ones presented in Section 1.1. The
differences in the calculations arise from the fact that we now take
into account that the X-ray gas density rises with decreasing radius
as $n \approxpropto r^{-1}$. Near the accretion radius it is typically
larger by a factor of few than at $\sim 1$ kpc (used by the above
authors).  Although the spatial resolution of the ROSAT HRI does not
allow us to directly probe the accretion radii in these objects
($0.03-0.05$ \kpc) the observed density profiles are very well fitted
by the power-law models (as detailed in Table~\ref{t:bh}) between
radii of $7-10\kpc$ down to $0.2 -0.5$ \kpc. The density estimates
indicate that accretion rates of order $0.5 - 2 \Msunpyr$ are
perfectly plausible in these systems.

The gas densities
at $R_{\rm A}$, estimated in Section 2, and the corresponding
accretion rates and best-fit model predictions indicate that the
systems are accreting at close to their upper limits implied by cooling. 
The main uncertainty in the calculation of $\mdot$ in these systems -
given the measured black hole masses - are the estimates of the
temperature of the cooling flows near the accretion radii. As
discussed by Gruzinov (1999; and along the same lines by Ostriker et
al. 1976) turbulent heat conduction may cause the flow to become
hotter and hinder accretion (although heat conduction is likely to be
suppressed in a magnetized cooling flows; see Chandran, Cowley \&
Sydora 1998). Alternatively, thermal instabilities may set up near the
sonic point and remove lots of the gas - although a central point mass
would eventually re-suppress the instability at small radii.  These
issues might be more easily addressed observationally in the near
future with high spatial resolution X-ray observations made with the
Chandra Observatory, which will determine the temperature profiles in
elliptical galaxies on spatial scales approaching the accretion
radius.

Given that the accretion rates in the ellipticals are similar to those
in Seyfert galaxies and other active nuclei, it is puzzling why these
objects should exhibit such different behavior. The different
environments in the galaxies may play a major role (see Fabian \& Rees
1995; Begelman 1986; Rees et al. 1982). When accretion proceeds from
the (already hot) interstellar medium (with relatively low angular
momentum) the resulting quasi-Bondi flow may go directly into a hot,
radiatively inefficient solution.
Such a solution might be preferentially obtained in elliptical
galaxies because the magnetic field strengths can be typically much
higher and close to equipartition values, even outside the accretion
radius (when the material has participated in a cooling flow; Soker \&
Sarazin; 1990). An initially weak magnetic field can be dramatically
increased by the radial inflow and shear in a cooling flow (or any
spherical inflow; Shapiro 1973; Meszaros 1975) - the combination of
transverse compression and radial compression can enhance the magnetic
stresses proportionally to $r^{-4}$, whereas the gas pressure goes no
more steeply than $r^{-5/2}$. The magnetic field is built up to
equipartition values even before it reaches the accretion radius.  If
the material enters the accretion flow with close to equipartition
magnetic fields, it might be more likely to accrete directly in a
high-$\alpha$ flow (as $\alpha$ is expected to scale with $\beta$;
e.g.  Hawley, Gammie \& Balbus 1996. Higher viscosity (if $\alpha
\approxgt 0.1$ can be achieved) would then allow the systems to be
radiatively inefficient.  The presence of relatively large scale
(mostly radial) magnetic fields might also be important as an agent
for launching the wind itself. Finally, since at large distances from
the disk the inertia of the gas can cause the magnetic field to become
increasingly toroidal, the magnetic stresses could also be responsible
for converting the centrifugal outflow into a more collimated jet
structure (as observed in the different degrees of radio activity in
most of the ellipticals).

\section{Alternatives to the Wind Model} 

\subsection{The Bremsstrahlung Interpretation}

It is worth being slightly more explicit about precisely what is
required to reconcile the basic ADAF model of Narayan \& Yi (1995b)
with the radio (DM99a) and X-ray observations (Paper I) of nearby
elliptical galaxies.

In particular, we have been arguing as if the bremsstrahlung
interpretation of the observed hard X-rays fixes the accretion rate to
be $\sim$ the Bondi rate.  This is not strictly correct; the
bremsstrahlung luminosity depends only on the gas density and
temperature, not on the accretion rate.  What the X-ray observations
fix, then, is $\rho(R_{\rm out}) \propto \dot M(R_{\rm
out})/|v_r(R_{\rm out})|$ where $v_R(R_{out})$ is the radial velocity
of the gas near the outer radius of the accretion flow. The Bondi
solution has $v_r \approx v_K$ while the ADAF solution has $v_r
\approx \alpha v_K$, where $v_K$ is the Keplerian or free fall
velocity at $R_A$.  Consequently, the X-ray observations fix $\dot
M(R_{\rm out})/\alpha$ (cf Narayan \& Yi 1995b).

The radio observations, on the other hand, are sensitive to
$\rho(R_{\rm in}$) and $T_e(R_{\rm in})$, the flow properties close to
the black hole.  Since $v_R \sim c$ near the black hole (gas flows
through the horizon at the speed of light), the radio observations
indeed fix $\dot M(R_{\rm in})$ (given $\alpha$, $\beta$, $\delta$,
etc. which fix $T_{\rm e}$).  In fact, they fix $\dot M(R_{\rm in})$ to be $\ll
\dot M_{\rm Bondi}$.

In the previous subsection, we emphasized that the accretion rates
necessary to explain the X-rays as bremsstrahlung are comparable to
the Bondi value.  This is, however, for $\alpha \approx 0.1$ (the
value chosen here).  If $\alpha$ were to be $\ll 1$, the accretion
rates consistent with the X-ray observations would also be $\ll \dot
M_{\rm Bondi}$.  Consistency with the observations would require
$\alpha \sim 10^{-4}-10^{-3}$.  In light of recent work on MHD
turbulence and angular momentum transport in accretion flows (Balbus
\& Hawley 1998; Hawley, Gammie, \& Balbus 1996), we do not think that
small $\alpha$ due to small viscosity is a viable possibility.  A more
interesting possibility is a suggestion due to Gruzinov (1999). He
pointed out that if there is significant radial heat conduction in a
Bondi-type flow, 
it can lead to a strong suppression in the radial velocity of the
accreting material near $R_A$.  Thus, even in the absence of viscosity
(Gruzinov's model was non-ideal but inviscid), the radial velocity may
be much smaller than the free fall value.  His model could, on these
grounds, account for the observations of nearby ellipticals as well as
the wind model we have been emphasizing (without any outflow).

There is, however, one argument which we believe rules out Gruzinov's
proposal as a complete explanation for the observations of DM99 and
Paper 1.  Namely, any model with $\alpha \ll 1$ which produces X-ray
luminosities comparable to those observed will have a cooling time
much shorter than the inflow time of the gas and will therefore cool
catastrophically to a thin disk.  More concretely, only for $\dot
m(R_{\rm out}) \approxlt \alpha^2$ can cooling be neglected in the
dynamics of the flow (as is done in the ADAF and Bondi solutions).
The bremsstrahlung interpretation of the X-rays from nearby
ellipticals fixes $\dot m(R_{\rm out})/\alpha$ to be $\sim 0.1$ (cf
Table 4 and recall that those models took $\alpha = 0.1$).
Consequently, only models with $\alpha \approxgt 0.1$ are consistent
hot accretion flow models for the systems at hand.  This rules out any
solution, including Gruzinov's, which appeals to a low radial velocity
near $R_A$. If indeed heat conduction (or other mechanisms) reduces
$\Mdot(R_{\rm out}) \ll \Mdot_{\rm Bondi}$ (but $\alpha$ is still
$\sim 0.1$) the observed X-ray flux (and the radio one, if the
decrease is very severe) must be produced by processes other than
bremsstrahlung.


\subsection{Photon Starved Comptonization and Non-thermal Emission}

In the above analysis we have assumed that the accreting plasma is
thermal and that there is no contribution to the observed emission
from non-thermal electrons. We have also assumed that any strong
winds/outflows are non-radiative (this can be shown to be a plausible
assumption within the context of a particular hydro-magnetic wind;
Blandford \& Payne 1982).  Nevertheless, dissipation (in a
flow/outflow with a shearing, near-equipartition, magnetic field) is
likely to occur and, at least in part, it will do so in transient
localized regions (behind shocks in the outflow or in sites of
reconnection both in the flow and outflow) where a small number of
electrons achieve ultra-relativistic energies. These electrons would
cool much more rapidly and their synchrotron emission would extend to
much higher energies (particularly if Comptonization is important)
than for thermal electrons.  The efficiency of these processes would
be much higher, implying (in order to reproduce the observed
low-luminosities) that either the accretion rates (i.e., the
densities) are much lower than expected from the estimates given in
Section 1, or that such processes accelerate only small number of
particles in localized regions.

The observed radio spectrum and, in particular, its extension to the
infrared (and possible higher frequencies), depends on how efficiently
non-thermal particles can be accelerated. The fairly sharp cut-offs
implied by the VLA and SCUBA observations (of the Virgo ellipticals
and NGC 1399 - but not M87) imply that any synchrotron emission from a
non-thermal distribution is unlikely to contribute significantly
beyond $\sim 10$ GHz and therefore will not extend to X-ray energies.

One possible non-thermal model for the observed emission is that the
same population of electrons produces both the radio/sub-mm emission
by synchrotron radiation and the X-ray emission by the inverse Compton
process.  In this case, we can describe the relativistic electrons by
a broken power-law distribution function, with spectral indices
smaller and larger than 3, below and above a break energy $\gamma_{b}$
(in units of $m_{\rm e} c^{2}$), respectively (e.g.  $N(\gamma)
\propto \gamma^{-n_{1,2}}$ with $n_1 \sim 1-1.5$ for $\gamma <
\gamma_b$ and $n_2 \sim 3-4$ for $\gamma > \gamma_b$).  The peak of
the synchrotron emission occurs at $ \nu_{\rm S} \approx (4/3)\nu_{\rm
  B}\gamma^2_{\rm b}$, where $\nu_{\rm B}=2.8 \times 10^6 B$ Hz is the
Larmor frequency.

The observed spectral energy distributions for our sources indicate
that the luminosity of the soft/synchrotron emission is typically a
few orders of magnitude less than the X-ray luminosity, implying that
the Compton spectra would arise from a photon starved plasma. The
situation envisaged here resembles the synchrotron-self Compton model
developed for BL Lac sources (e.g., Ghisellini 1989; Ghisellini,
Maraschi \& Dondi 1996; Ghisellini et al.  1998; in BL Lacs, Compton
scattering of other soft photons, such as those arising from a
standard accretion disk or the broad line region, can also be
important; neither of these are, however, thought to be present in the
elliptical galaxies of interest).\footnote{One crucial difference,
  however, is the absence of a radio cutoff in the spectrum of BL
  Lacs.  This has important implications for the inferred parameters of
  nonthermal models, as is discussed below.}

In those models (and others developed within the context of
non-thermal photon starved - electron or pair - plasmas; Zdziarski \&
Lamb 1986; Zdziarski, Coppi \& Lamb 1990) the sources are very compact
(where the compactness parameter is defined by $\ell= L/R
(\sigma_{T}/mc^3)$) which implies that all of the particles cool
before escaping the source ($t_{\rm cool} \sim \pi t_{\rm cross}[\ell
(\gamma -1)]^{-1}$ and $t_{\rm cross} \sim R/c$). A steady state can
therefore be achieved.

A steep particle distribution is continuously injected; due to
self-absorption, synchrotron cannot emit the bulk of this luminosity;
in this case, it is primarily produced by multiple Compton scatterings
(and flat X-ray energy indices $<< 1$ can be obtained). For the
sources in ellipticals, however, $\ell <1$ even when $r=1$. It is
therefore difficult to have an optical depth close to unity and for
multiple Compton scatterings to be important.

The injected power in non-thermal electrons can nonetheless still be
greater than the synchrotron luminosity and would emerge as first
order Compton scattering peaking at X-ray wavelengths.  The maximum
power in Inverse Compton emission will occur at a frequency $\nu_{\rm
C} \approx (4/3) \gamma_{b}^2 \nu_{\rm S}$ where, from observations,
$\nu_{\rm C} \approxgt 10 \keV$. Given the frequency of the radio peak
and a lower limit for $\nu_{\rm C}$ we can derive the energy of the
electrons contributing most of the power and the required magnetic
field strength: $\gamma_{\rm b} \approx \left(3\nu_{\rm C}/[4\nu_{\rm
S}] \right)^{1/2} \sim 10^4$ and $B \sim 10^{-5}$ Gauss.

The required magnetic field strength $\sim 10^{-5}$ Gauss is an
implausibly small fraction of the equipartition value for the
accretion flow (even in the presence of a strong wind, $B_{\rm equi}
\sim 1$ Gauss near $r \sim 1$).  Thus, it is unlikely that nonthermal
synchrotron self-Compton processes in the accretion flow are important
contributors to the observed emission.  It is also not likely that
field strengths as low as $\sim 10^{-5}$ Gauss could be achieved in an
outflow/wind or jet. Equipartition with the radiation energy density
(in particular for localized regions) would still predict much higher
field strengths. In any case, for such photon starved cases
bremsstrahlung emission from both a thermal or non-thermal population
is likely to be important.  If the X-ray emission were due to e.g.
second (or third) order Compton scattering (for a similarly peaked
electron distribution but with $\gamma_{\rm b} \sim$ a few 100), the
limits on $B$ would be relaxed. In order for the the second order
Compton scattering to contain the required X-ray luminosity, $\tau$
would need to be high (e.g. $\sim 1$, since the probability for a
second order Compton scattering scales as $\tau^2$). Given the
observational constraint that $\ell \ll 1$, large enough optical depths
would seem difficult to achieve (given the highly sub-Eddington
nature of the sources the 'jet' emitting region needs to be localized
and have low density).


We note that the case is different for M87. For this object a sharp
spectral turnover in the radio band is not required by the data (see
Fig.~\ref{spec}a). The spectral energy distribution (radio, optical
data and the hard X-ray spectra) is still somewhat double peaked but
with similar luminosities in both the radio-optical and X-ray peaks.
From above, for $\gamma_{\rm b} \sim 100$ (and assuming a lower limit
for the high energy peak to be around $10 \keV$, although
contributions up to gamma-ray frequencies are likely to occur), the
magnetic field can be much larger than in the previous cases,
comparable to the equipartition value.  We cannot, therefore, exclude
 some relevant contribution to the spectral energy distribution
from non-thermal particles in M87. Note, although, that the highly
sub-Eddington luminosity of this source still puts very
strong constraints on the density and emitting region size. 
  
In conclusion, simple estimates of non-thermal synchrotron
self-Compton models for the observed radio and X-ray emission in
nearby ellipticals suggest that highly sub-equipartition magnetic
fields are required ($B \sim 10^{-5}$ Gauss).  This is contrary to
expectations, both for the accretion flow and for a magnetically
driven wind or a collimated outflow.  We cannot, however, rule out
other nonthermal models for the observed emission such as a jet seen
off--axis with separate populations of nonthermal electrons producing
the observed radio and X-ray emission by synchrotron radiation, or an
unobserved source of soft photons in the optical/IR available for
Compton scattering, etc. Although these might quickly become rather
contrived.  

The most promising means of distinguishing bremsstrahlung and
non-thermal models is likely to be variability, which is discussed in
the next section.

\section{Variability}

Here we give a brief discussion of the variability properties of the
bremsstrahlung model, the salient feature of which is that {\em there
should be little to no short timescale ($\sim$ day to month) X-ray
variability because the bremsstrahlung emission arises from relatively
large radii in the accretion flow}.

The bremsstrahlung emission from a radius $r$ in the accretion flow
can be expected to vary by (at most) order unity on a timescale
comparable to the local dynamical time, given by $t_d \approx
(R^3/GM)^{1/2} \approx 4 m_9 r^{3/2}$ hours.\footnote{For
  bremsstrahlung, it is the dynamical time, not the light crossing
  time, which is relevant.}  For timescales less than $t_d(r)$ the
variability from emission at radius $r$ is suppressed.  An upper limit
to the suppression can be made by noting that for $t' < t_d(r)$, there
can still be (at most) order unity variability from fluctuations on
length scales $r' \approx v_K(r) t'$.  There are $\sim (r/r')^3$ such
blobs varying incoherently in a shell of radius $r$ and thickness $dr
\sim r$.  Consequently, for $t' < t_d(r)$, the variability amplitude
should be $< [t'/t_d(r)]^{3/2}$.  This is a strict upper limit because
in reality there will be a power spectrum of fluctuations with small
scale fluctuations having less power than large scale fluctuations.

The above analysis implies that the fractional luminosity fluctuation
at frequency $\nu$ on a timescale $\delta t$ is given by \beq {\delta
L_\nu(\delta t) \over L_\nu} = {\int d\log r \ \epsilon_\nu(r) \
W(r,\delta t) \over \int d\log r \ \epsilon_\nu(r)}, \label{var} \eeq
where $\epsilon_\nu$ is the bremsstrahlung emissivity (examples of
which are in Figure~\ref{bremfig}) and \beq W(r,\delta t) = A \ {\rm
Min} \left(1, \left[{\delta t \over t_d(r)} \right]^{3/2} \right)
\label{window} \eeq is the power spectrum for the particular model
given above ($A \sim 1$ is the amplitude of the fluctuation on the
dynamical time).\footnote{If, as they should be, the bremsstrahlung
emissivity fluctuations are due to density fluctuations, the $3/2$
power law in equation (\ref{window}) becomes $3/2 + 2/3$ (for
Kolmogorov turbulence, either hydrodynamic or magneto-hydrodynamic).}

  A crude approximation to the above integral can be obtained by
  setting $\epsilon_\nu(r) \propto \delta (r - r_\nu)$, where $r_\nu$
  is the radius near which most of the emission at frequency $\nu$
  originates.  In this case, \beq{\delta L_\nu(\delta t) \over L_\nu}
  \sim A \ {\rm Min} \left(1, \left[{\delta t \over t_d(r_\nu)}
  \right]^{3/2} \right). \label{simpvar} \eeq
  
  Even in the absence of a wind, $r_\nu \approxgt 100$ for all X-ray
  energies of interest (see Fig.~\ref{bremfig}).  In fact, in the soft
  X-rays ($\sim 1$ keV), $r_\nu \approx 10^3-10^4$.  For wind models,
  the value of $r_\nu$ in the soft X-rays is unchanged while in the
  hard X-rays ($\approxgt 10$ keV) it becomes $r_\nu \approx r_{\rm
  out}$.  That is, the bremsstrahlung emission in the hard X-rays is
  dominated by the radius at which the wind becomes important (see
  Fig.~\ref{bremfig}b).  Consequently, variability studies at
  $\approxgt 10$ keV can explicitly determine the value of $r_{\rm
  out}$.
  
  For $r_\nu \sim 100$, the dynamical time is $\approx$ 1 year.  It is
  clear, then, that within the bremsstrahlung interpretation there
  should be no variability in the observed X-rays over a single
  observing run.  Furthermore, multiple observing runs should detect
  significant variability only if separated by months or even years.

  To be slightly more explicit, Figure~\ref{varfig} shows the expected
  variability (calculated from eq. [\ref{var}]) for five X-ray
  energies for the model whose emissivity curves are shown by dotted
  lines in Figure 1, namely a model with a wind with $r_{\rm out} =
  10^3$ and $p = 0.5$ (recall that these are observationally favored
  parameters).  This model assumes that only bremsstrahlung
  contributes to the X-ray emission (no Comptonization or nonthermal
  particle emission).  Note that the relevant dimensionless
  variability timescale is $\delta t/t_d(r) \propto \delta t/m_9$
  (hence the abscissa in Fig.~\ref{varfig}).
  
  For $\delta t \sim$ a day, the luminosity fluctuation expected is
  $\approxlt 0.1 \%$.  Even for $\delta t \sim$ a month, $\delta
  L_\nu/L_\nu \approxlt 10 \%$ around $100$ keV and $\delta
  L_\nu/L_\nu \approxlt 0.3 \%$ around $1$ keV.  Only on a timescale
  of $\sim$ a year do we expect significant X-ray variability.  Note
  also the clear trend with X-ray energy: the variability is greater
  at higher X-ray energies since this emission arises from closer to
  the black hole. 

Observations of short timescale variability would imply that either the
flow is highly inhomogeneous and bremsstrahlung emission is produced in
localized, higher density regions or that non-thermal SSC of thermal
Comptonization in the central region is producing the observed flux.

One further diagnostic for  testing the bremsstrahlung hypothesis
is the observations of thermal X-ray lines, a test that should be
plausibly carried out with the Chandra Observatory (Narayan \& Raymond
1999). If the X-ray emission is dominated by the bremsstrahlung
emission the X-ray spectrum of the accretion flow should exhibit
thermal X-ray lines (Narayan \& Raymond 1999). Their presence would
also be an important discriminant for assessing the outer radius of
the flow (because of the strong temperature dependence and the
one-to-one mapping between temperature and radius the bulk of the line
emission is expected to originate from radii outside $10^{4}$; Narayan
\& Raymond 1999).

\subsection{M87}

\begin{figure}
\centerline{
\vbox{
\psfig{figure=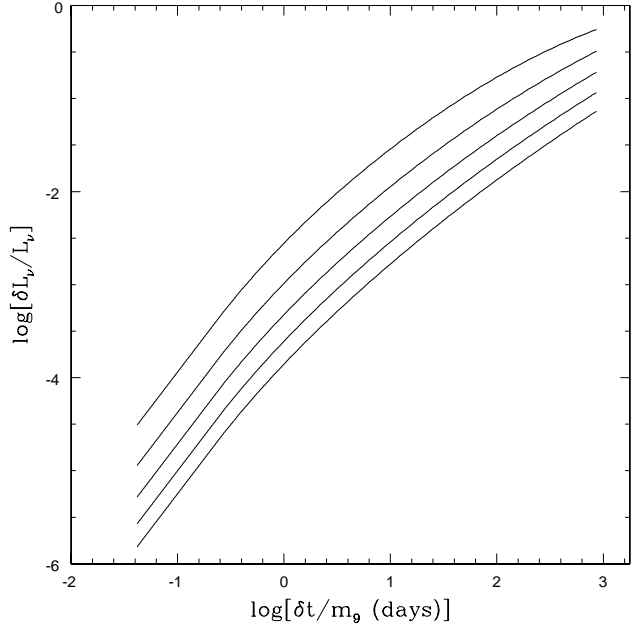,width=0.5\textwidth}
}}
\caption{The fractional variability expected at five X-ray energies 
  ($100, 30, 10, 3\ \& \ 1$ keV, from top to bottom) as a function
  of the timescale of observation in days.  The figure assumes pure
  bremsstrahlung emission in the X-ray band and a spectral model with
  $p = 0.5$ and $r_{\rm out} = 10^3$.}
\label{varfig}
\end{figure}

{\it ROSAT} HRI observations ($\sim 1$ keV) of the ``core'' of M87
show $\approx 20 \%$ variability on timescales of $\approx 6$ months
to a year (Harris, Biretta, \& Junor 1997).  This is not easy to
reconcile with a wind-dominated bremsstrahlung model for the X-ray
emission.  For M87, $m_9 \approx 3$.  Thus, if a model with $p \approx
0.5$ and $r_{\rm out} \approx 10^3$ is appropriate for M87,
bremsstrahlung alone would lead to $\approxlt 1\%$ variability at
$\approx 1$ keV on timescales of $\approx$ a year (cf
Fig.~\ref{varfig}), inconsistent with the observations.

One possible resolution of this discrepancy is that $r_{\rm out}$ is
$\ll 10^3$ so that there is significant X-ray emission from closer to
the black hole.  We find that if $r_{\rm out} \approxlt 30$, $\sim 10
\%$ soft X-ray variability can be produced (note from
Fig.~\ref{bremfig}a that there is a $\sim 10 \%$ contribution to the
soft X-ray bremsstrahlung luminosity from $r \sim 10$ in the absence
of a wind).  If such a small value of $r_{\rm out}$ is applicable to
all of the members of our sample, however, very large values of $p
\approxgt 1$ are required to reproduce the observed X-ray to radio
luminosity ratios (note that this is {\it not} required for M87). 

As emphasized in previous sections, however, M87 is a peculiar member
of our sample.  It is the only system for which high frequency radio
observations do not show a strong suppression (and therefore there are
no strong arguments against significant emission from nonthermal
particles; see the previous section).  Its one sided jet, likely to be
pointing towards us, is known to contribute to the emission in the
very core region (Baath et al. 1992) and to produce X-rays (knot A has
a slightly smaller X-ray luminosity than the core; Harris et al. 1997)

The ``core emission'' in this case is likely to contain significant
contamination from unresolved jet emission, in addition to accretion
flow emission.  This is likely the origin of the variable soft X-ray
emission.  Two reasons argue against thinking that all of the X-ray
emission is due to unresolved jet emission with a negligible accretion
flow contribution. (1) Accretion at the Bondi rate in M87 should
produce an X-ray luminosity comparable to that observed and (2) there
is a clear correlation between the hard power law X-ray emission and
the host galaxy X-ray emission in the six ellipticals we are studying
(cf Fig. 3 of Paper 1 and discussion therein).  Excluding M87, these
systems are not believed to have jets which can produce hard X-rays at
the level observed.  Since M87 nicely satisfies the observed
correlation, the non-jet emission should therefore be comparable to
the total emission.

For a number of reasons, M87 is therefore a poor system in which to
probe the (lack of) variability expected from bremsstrahlung emission in
our models.  More promising candidates are the Virgo ellipticals
(NGC 4472, 4636, and 4649).  The large X-ray/radio luminosity ratios
in these systems (in contrast to M87) imply negligible contributions
from Comptonized synchrotron photons in the X-ray band.  Jet emission
in X-rays is also expected to be small in these systems.  Barring
nonthermal particle emission from the accretion flow/wind (see \S7),
the X-ray emission should therefore be dominated by bremsstrahlung
processes, providing an excellent test of our variability predictions.

\section{Summary and discussion}
The discovery of hard power-law emission (Paper I) from a sample of
elliptical galaxies has brought to a sharper focus the study of nearby
supermassive quiescent black holes. We have discussed how the broad
band spectral energy distributions of this sample of elliptical
galaxies, accreting from their hot gaseous halos at rates comparable
to their Bondi rates, can be explained by low-radiative efficiency
accretion flows in which a significant fraction of the mass, angular
momentum and energy is removed from the flows by winds. Such outflows
are not simply an added parameter in the model but a dynamically
important component of accretion at low radiative efficiency
(Blandford \& Begelman 1999). The characteristically suppressed
synchrotron emission in the radio band (excluding M87) and the
systematically hard X-ray spectra, as expected from thermal
bremsstrahlung processes, strongly supports the conjecture that
significant mass outflow is a natural consequence of systems accreting
at low-radiative efficiencies.

Bremsstrahlung emission in the $2-10$~\keV~band is produced primarily
at large radii in the accretion flow and is sensitive to the rate at
which matter is fed to the flow ($\dot m_{\rm out}$). By contrast,
synchrotron emission originates from the interior regions and, for a
given value of $\mdot_{\rm out}$, decreases significantly with
increasing $p$ (where $p$ is the power-law index of the accretion rate
in the flow, i.e., $\mdot \propto \mdot_{\rm out}(r/r_{\rm out}^p$)
for $r < r_{\rm out}$).  We have shown that radio observations (DM99)
and ASCA X-ray fluxes can be explained by ADAF models only if
$\mdot_{\rm out}$ is of order the Bondi value $\sim 0.001 - 0.02$,
$p\sim 0.5-1$ and $r_{\rm out} \sim 100 -1000$.  Significant electron
heating (e.g. $\delta \approxgt 0.1$) is also fully consistent with
the outflow models for ellipticals. In view of this, Sgr$^*$ A might
not be very different from the elliptical nuclei we have studied here
and can also be explained within the context of strong mass loss
models (with the strict requirement, in order to satisfy the firm VLBI
constraints, that $\delta \approxgt 0.1$ in the Galactic centre; see
QN99).

The small values of $r_{\rm out}$ relative to the accretion radius
($r_A \sim 10^6$) may suggest that angular momentum only becomes
important well inside the accretion radius.

We have examined possible contributions from non-thermal particle
distributions (which are likely to originate in shock regions and
reconnection sites in the flow/outflows, or in the observed jets). The
observed lack of any extension of the synchrotron component towards IR
and optical wavelengths, and the typically high ratio of X-ray to
radio luminosity (in the Virgo ellipticals and NGC1399), places strong
limits on the magnetic fields strengths, implying that any non-thermal
components are unlikely to contribute significantly to the observed
emission.  A bremsstrahlung component is more likely to dominate the
X-ray emission than SSC. The case of M87 is less clear and non-thermal
contributions, both at low and high energies, may occur. However, the
highly sub-Eddington luminosity of M87 (and the other galaxies in our
sample) does not allow a simple connection with jet dominated, Blazar-like
sources to be made.  Theoretical models and observations at higher energies
(e.g. gamma rays) are required to better assess the importance of non-thermal
models. Outflows, if present, should contain most of the energy
and mass but need to be very radiatively inefficient.

The predicted absence of short timescale variability in the X-ray band,
which is expected if the X-ray emission is primarily bremsstrahlung 
emission (which in wind models originate from $r_{\rm out} \sim 100-1000$, where the 
dynamical timescale is months to years) will be readily assessed with 
observations made with the Chandra observatory. If the
X-ray emission is dominated by bremsstrahlung processes, the X-ray
spectrum may also exhibit thermal line emission (Narayan \& Raymond
1999). Their presence would also be an important discriminant for
assessing the outer radius of the flow (because of the strong
temperature dependence and the one-to-one mapping between temperature
and radius, the bulk of the line emission is expected to originate from
radii outside $10^{4}$; Narayan \& Raymond 1999).

As discussed in Paper I, even from a purely observational point of
view, the low luminosities of the sources we are considering and their
characteristically hard and energetically dominant X-ray spectra,
identify the elliptical galaxy nuclei as a new class of accreting
black holes, which can be clearly distinguished from Seyfert nuclei
(which are most often hosted in spiral galaxies). We suggest that the
difference in radiative efficiency (readily assessed in ellipticals
given the black hole mass and estimates of accretion rates),
manifested by the accretion flows close to the black holes, in these
different classes of objects, does not solely arise from a difference in
$\mdot$ (the Bondi rates are adequate to fuel an active nucleus in the
elliptical galaxies and a thin disk solution would be viable). 

The accretion solution adopted is likely to depend on the manner in
which material is fed into the nucleus (e.g. Rees et al. 1982;
Begelman 1986). In spiral galaxies, the bulk of the interstellar
medium resides in a disk and it is plausible that a radiatively
efficient flow persists all the way into the black hole. In elliptical
galaxies, most of the gas participating in the angular
momentum-dominated accretion flow originates from the hot interstellar
medium that pervades the galaxies (often forming a cooling flow). We
have speculated that the increase in magnetic field strength due to
the radial inflow and shear in a cooling flow (with the magnetic field
pressure in equipartition with thermal pressure within radii $\sim 10$
kpc) might be the primary cause for the ensuing high $\alpha$
accretion flow. (Material fed into the accretion flow would be highly
magnetized and $\alpha$ should scale $\propto$ to magnetic field
strength; e.g. Hawley et al. 1996). High viscosity parameters (if
$\alpha \sim 0.1$ can be achieved) would in turn give rise to
low-radiative efficiencies. The presence of relatively strong magnetic
fields in these environments can also play an important role for
driving and possibly collimating the outflows.  We suggest that the
outflows could be energized by loops of field anchored to the flow
itself (and responsible for driving the wind). At large distances from
the disk, the inertia of the gas can cause the magnetic field to
become increasingly toroidal. Magnetic stresses could be responsible
for converting the centrifugal outflow into a more collimated jet
structure. The development of radio structures in ellipticals may also
be fostered by the presence of a hot interstellar medium (and the more
prominent radio structures in central cluster galaxies - the FRI type
sources in the sample - are the ones to be found in the gas richer,
higher pressure environments; see also Fig. 3 in Paper I) and
partially suppressed by its absence in spirals.  We note that, if not
partially collimated, an outflow (which would contain most of the
accreted mass, $\sim 90 - 97$ per cent) could stifle the accretion
flow. If accretion is stifled the radio and X-ray emission are likely
to be produced by small numbers of non-thermal particles in shock
sites in the jets/outflows.

We further argue that low-radiative efficiency accretion and its
associated outflows maybe relevant for understanding radio-loud
AGN. We are suggesting that the more active M87, a classical FRI
source (see also Reynolds et al. 1996) together with NGC 1399 and NGC
4696 (also weak FRI sources) provide us with some of the strongest
evidence for low-efficiency accretion. X-ray emission
from FR-II sources is often associated with the presence of broad iron
$K\alpha$ fluorescence lines and therefore thin accretion disks
(e.g. 3C 109, Allen et al. 1996; 3C390.3 Eracleous, Halpern \& Livio
1996). The difference in accretion mode (e.g. see also Begelman 1985)
may also be manifested in the different
properties of FRI and FR-II sources.

The differences between elliptical and spiral nuclei does not only
arise from their different environments but also from their respective
histories: elliptical galaxies have black hole masses of $10^9$
-$10^{10}$\Msun, consistent with those expected if these galaxies have
undergone a quasar phase in the past (e.g. Salucci et al. 1998).
Black holes in spiral galaxies do not exceed $10^{8} \Msun$,
supporting the suggestion that accretion at low-radiative efficiencies
might be relevant in the final stages of accretion in early type
galaxies.

\section*{Acknowledgments}
We thank Chris Carilli for the high frequency VLA data of
NGC 1399 and Dimitrios Psaltis for very useful conversations. TDM
acknowledges support for this work provided by NASA through Chandra
Postdoctoral Fellowship grant number PF8-10005 awarded by the Chandra
Science Center, which is operated by the Smithsonian Astrophysical
Observatory for NASA under contract NAS8-39073. E.Q. is supported by
NSF Graduate Research Fellowship.


\begin{thebibliography}{}
\bibitem{} Allen S.W., Di Matteo T., Fabian A.C., 1999, MNRAS, submitted (Paper 1)
\bibitem{} Allen S.W., Fabian A.C., Idesawa E., Inoue H., Kii T., Otani C., 1997, MNRAS, 286, 765
\bibitem{} Abramowicz M., Chen X., Kato S., Lasota J.~P., Regev O., 1995, ApJ, 438, L37
\bibitem{} Balbus, S. A., Hawley, J. F., in Accretion Processes in Astrophysical Systems: Some Like it Hot! Eighth Astrophysics Conference, College Park, MD, October 1997. Edited by Stephen S. Holt and Timothy R. Kallman, AIP Conference
Proceedings 431., p.79
\bibitem{} Begelman M.~C., 1985, in Astrophysics of active galaxies and quasi-stellar objects, eds Miller J.~S., University Science Books, Mill Valley, P411
\bibitem{} Begelman M.~C., 1986, Nat, 322, 614
\bibitem{} Bisnovatyi-Kogan, G.S., Lovelace, R.V.E.,  1997, ApJ, 486, L43
\bibitem{} Blackman E.G., 1999, MNRAS, 302, 723
\bibitem{} Blandford R.D., Begelman M.C., 1999, MNRAS, 303, L1
 \bibitem{} Blandford R.D., Payne D.G., 1982, MNRAS, 199, 883 
\bibitem{} Bondi H., 1952, MNRAS, 112, 195
\bibitem{} Byun Y., et al. 1996, AJ, 111, 1889
\bibitem{} Chandran B., Cowley S., Sydora R., 1997, AAS, 191,5307 
\bibitem{} Di Matteo T., Fabian A.C., 1997, MNRAS, 286, L50
\bibitem{} Di Matteo T., Fabian A.C., 1997a, MNRAS, 286, 393
\bibitem{} Di Matteo T., Fabian A.C., Rees M.J., Carilli C.L., Ivison R.J. 1999, MNRAS, in press 
\bibitem{} Eracleous M., Halpern J.P., Livio M., 1996, ApJ, 459, 89
\bibitem{} Esin A.A., McClintock J.E., Narayan R., 1997, ApJ, 489, 865
\bibitem{} Fabian A.~C., Canizares C.~R., 1988, Nat, 333, 829
\bibitem{} Fabian A.~C., Rees M.~J., 1995, MNRAS, 277, L55
\bibitem{} Fabbiano G., Kim D.W., Trinchieri G., ApJS, 1992, 80, 531
\bibitem{} Ford H.~C. et al. 1995, ApJ, 1994, 435, L27
\bibitem{} Ghisellini G., 1989, MNRAS, 236, 341
\bibitem{} Ghisellini G., Maraschi L., Dondi L., 1996, A\&A Suppl. S., 120, 503
\bibitem{} Ghisellini G., Celotti A., Fossati G., Maraschi L., Comastri A., 1998, MNRAS, 301, 451
\bibitem{} Gruzinov A. V., 1999, ApJ, submitted
\bibitem{} Gruzinov A. V., 1998, ApJ, 1998, ApJ, 501, 787
\bibitem{} Harms R.~J. et al., 1994, ApJ, 435, L35
\bibitem{} Hawley J. F.,Gammie C. F.,Balbus, S. A., 1996, 464, 690
\bibitem{} Ho, L.~C. 1998, in Observational Evidence for Black Holes in the Universe, ed. S.~K. Chakrabarti (Dordrecht: Kluwer), 157
\bibitem{} Hummel E., Kotani C.~G., Ekers R.~D., 1983, A\&A, 127, 205 
\bibitem{} Kormendy J., Richstone D. 1995, ARA\&A, 33, 581 
\bibitem {} Macchetto, F., Marconi A., Axon D.J., Capetti A.,
 Sparks W., Crane P., 1997, ApJ, 489, 579
\bibitem{} Magorrian J. et al., 1998, AJ, 115, 2285 
\bibitem{} Mahadevan R., 1997, ApJ, 477, 585 
\bibitem{} McKee, C.F., Cowie L.L., 1977, ApJ, 215, 213
\bibitem{} Meszaros P., A\&A, 44, 59
\bibitem{} Nakamura K.E., Kusunose M., Matsumoto R., Kato S., 1997, PASJ,49,503
\bibitem{} Narayan R., Yi I., 1994, ApJ, 428, L13 
\bibitem{} Narayan R., Yi I., 1995a, ApJ, 444, 231
bibitem{} Narayan R., Yi I., 1995b, ApJ, 452, 710
\bibitem{} Narayan R., Raymond J., 1999, ApJ, 515, L69
\bibitem{} Narayan R., Barret D., McClintock J., 1997, ApJ, 482, 448
\bibitem{} Narayan R., Mahadevan R., Quataert E., 1998, to appear in "The Theory
 of Black Hole Accretion Discs", eds. M. A. Abramowicz, G. Bjornsson, and
J. E. Pringle, (Cambridge University Press)
\bibitem{} Nulsen P.E.J., Stewart G.C., Fabian A.C., 1984, 208, 185
\bibitem{} Quataert E., 1998, ApJ, 500, 978
\bibitem{} Quataert E., Gruzinov A.V., 1999, ApJ, in press
\bibitem{} Quataert E., Narayan R., 1999, ApJ, in press
\bibitem{} Rees M.~J., 1982, in Riegler G., Blandford R., eds, The Galactic Center. Am. Inst. Phys., New York, 166
\bibitem{} Rees M.J., 1987, MNRAS, 228, 47p
\bibitem{} Rees M.~J., Begelman M.~C., Blandford R.~D., Phinney E.~S., 1982, NAT., 295, 17 
\bibitem{} Reynolds C.~S., Di Matteo T., Fabian A.~C., Hwang U., Canizares C.~R., 1997, MNRAS, 283, L111 
\bibitem{} Sadler E.~M., Jenkins C.~R., Kotanji C.~G., 1989, MNRAS, 240, 591
\bibitem{} Salucci P., Szuszkiewicz E., Monaco P., Danese L., 1998, MNRAS submitted
\bibitem{} Shapiro S.L., 1973, ApJ, 185, 69
\bibitem{} Slee O.~B., Sadler E.~M., Reynolds J.~E., Ekers R.~D., 1994, MNRAS, 269, 928
\bibitem{} Soker N., Sarazin C., 1990, ApJ, 348, 73
\bibitem{} Stewart G.C., Canizares C.R., Fabian A.C., Nulsen P.E.J., ApJ, 278, 536
\bibitem{} Wrobel J.~M., 1991, AJ, 101, 127
\bibitem{} Wrobel J.M., Heeshen D.S., 1991, AJ, 101, 148
\bibitem{} Van der Marel R.P., 1991, MNRAS, 253, 710
\bibitem{} van der Marel R.P., 1998, ApJ, submitted, (astro-ph/9806365)
\bibitem{} Zdziarski A., Lamb D.Q., 1986, 309, L79
\bibitem{} Zdziarski A., Coppi P.S., Lamb D.Q., 1990, ApJ, 357, 149

\end{thebibliography}
\end{document}